\documentclass[12pt]{article}
\usepackage{latexsym,amsfonts}
\usepackage[dvips]{graphicx}
\def\a{\alpha}
\def\b{\beta}
\def\g{\gamma}
\def\d{\delta}
\def\e{\eta}

\def\l{\lambda}

\def\m{\mu}
\def\n{\nu}
\def\r{\rho}
\def\o{\omega}
\def\s{\sigma}
\def\S{\Sigma}

\def\p{\pi}
\def\e{\varepsilon}
\def\cA{{\cal A}}

\def\be{\begin{equation}}
\def\ee{\end{equation}}
\def\beq{\begin{eqnarray}}
\def\eeq{\end{eqnarray}}
\def\nn{\nonumber}
\def\ca{{\cal A}}

\def\ce{{\cal E}}
\def\cf{{\cal F}}

\def\ch{{\cal H}}

\def\ck{{\cal K}}

\def\cm{{\cal M}}
\def\cn{{\cal N}}

\def\cv{{\cal V}}

\def\RR{{\mathbb{R}}}

\def\ZZ{{\mathbb{Z}}}
\newcommand{\ft}[2]{{\textstyle {\frac{#1}{#2}} }}

\newcommand{\frakh}{{\mathfrak{h}}}

\newcommand{\Rn}{{\mathbb{R}}}
\newcommand{\E}{E_{10}}

\newcommand{\Ref}[1]{(\ref{#1})}
\newcommand{\non}{\nonumber\\}

\newcommand{\bqn}{\begin{eqnarray}}\newcommand{\eqn}{\end{eqnarray}}

\title{Cosmological Singularities, Einstein Billiards and Lorentzian Kac-Moody Algebras}
\author{Thibault Damour}
\date{\it Institut des Hautes \'Etudes Scientifiques, 35 route de Chartres, 
91440 Bures-sur-Yvette, France}

\begin{document}

\maketitle

\begin{abstract}
The structure of  the general, inhomogeneous solution of (bosonic) Einstein-matter
systems in the vicinity of a cosmological
singularity is considered. We review the proof (based on ideas of Belinskii-Khalatnikov-Lifshitz and
technically  simplified by the use of the Arnowitt-Deser-Misner Hamiltonian
formalism) that the asymptotic behaviour, as one approaches the singularity,
of the general solution is  describable, at each (generic)
spatial point, as a billiard motion in an auxiliary Lorentzian space.
For certain Einstein-matter systems, notably for pure Einstein gravity in
any spacetime dimension $D$ and for the particular Einstein-matter systems
arising in String theory, the billiard tables describing asymptotic
cosmological behaviour are found to be identical to the Weyl chambers of
some Lorentzian Kac-Moody algebras. In the case of the bosonic sector of
supergravity in 11 dimensional spacetime the underlying Lorentzian algebra
is that of the hyperbolic Kac-Moody group $E_{10}$, and there exists some
evidence of a correspondence between the general solution of the
Einstein-three-form system and a null geodesic in the infinite dimensional
coset space $E_{10} / K (E_{10})$, where $K (E_{10})$ is the maximal
compact subgroup of $E_{10}$.

\end{abstract}

\section{Introduction and general overview}
\setcounter{equation}{0}
\setcounter{theorem}{0}
\setcounter{lemma}{0}

A series of works \cite{DH1,DH2,DH3,DHJN,DdBHS,DHN2,Damour:2002et} 
has uncovered a remarkable connection between the asymptotic behaviour, 
near a cosmological singularity, of certain Einstein-matter systems and 
billiard motions in the Weyl chambers of some corresponding Lorentzian 
Kac-Moody algebras. This simultaneous appearance of {\it billiards}
(with {\it chaotic} properties in important physical cases) and of an underlying
{\it symmetry} structure (infinite-dimensional Lie algebra) is an interesting
fact, which deserves to be studied in depth. Before explaining in detail the
techniques (including the Arnowitt-Deser-Misner Hamiltonian formalism \cite{ADM})
that have been used to uncover this fact, we shall start by reviewing
 previous related works, and by stating the main results of this billiard/symmetry
 connection.
 
The simplest example of this connection concerns the
pure Einstein system, {\it i.e.} Ricci-flat spacetimes. Long ago, Belinskii,
Khalatnikov and Lifshitz (BKL) \cite{BKL1,BKL2,BKL3} gave a description of 
the asymptotic behaviour, near a spacelike singularity, of the general 
solution of Ricci tensor $=0$ in $(3+1)$-dimensional spacetime in terms of a
continuous collection of second-order, non linear ordinary differential 
equations (with respect to the time variable). As argued by these authors,
near the singularity the spatial points $x^i$, $i = 1,2,3$, essentially 
decouple, and this decoupling allows one to approximate a system of {\it 
partial} differential equations (PDE's) in 4 variables $(t,x^i)$ by a 
3-dimensional family, parametrized by $(x^i) \in {\mathbb R}^3$, of {\it 
ordinary} differential equations (ODE's) with respect to the time variable
$t$. The coefficients entering the nonlinear terms of these ODE's depend on 
the spatial point $x^i$ but are the same, at each given $x^i$, as those 
that arise in some spatially homogeneous models. In the presently 
considered 4-dimensional ``vacuum'' ({\it i.e.} Ricci-flat) case, the
spatially homogeneous models that capture the behaviour of the general 
solution are of the Bianchi type IX or VIII (with homogeneity groups ${\rm 
SU} (2)$ or ${\rm SL} (2,{\mathbb R})$, respectively). The asymptotic 
evolution of the metric was then found to be describable in terms of a 
chaotic \cite{KLL,Bar} sequence of generalized Kasner ({\it i.e.} 
power-law) solutions, exhibiting ``oscillations'' of the scale factors 
along independent spatial directions \cite{BKL1,BKL2,BKL3}, or, 
equivalently, as a billiard motion \cite{Chitre,Misnerb} on the 
Lobachevskii plane.

This picture, found by BKL in the 4-dimensional pure Einstein case, was then 
generalized in various directions: namely by adding more spatial 
dimensions, or by adding matter fields. Remarkably, it was found that there 
exists a {\it critical} spacetime dimension for the asymptotic behaviour of 
the general Ricci-flat solution \cite{DHS}. When $D \leq 10$ the solution
exhibits a BKL-type never-ending oscillatory behaviour with strong chaotic 
properties, while when $D \geq 11$ the general solution ceases to exhibit 
chaotic features, but is instead asymptotically characterized by a 
monotonic Kasner-like solution \cite{DHS,DHRW}. The addition of matter 
fields was also found to feature a similar subcritical/overcritical 
classification \cite{BK1,DH1,DHRW}. Let us consider, in some given 
spacetime dimension $D$, the Einstein-dilaton-$p$-form system, {\it i.e.} 
let us add a massless scalar $\phi$, and one or several $p$-form fields $A 
= A_{\mu_1 \ldots \mu_p} \, dx^{\mu_1} \wedge \ldots \wedge dx^{\mu_p} / 
p!$. It was found that the general solution, near a cosmological 
singularity, of this system is monotonic and Kasner-like if the dilaton 
couplings $ \lambda_p $ (defined below)
 of the $p$-forms belong to some (dimension dependent) open 
neighbourhood of zero \cite{DHRW}, while it is chaotic and BKL-like if the 
dilaton couplings belong to the (closed) complement of the latter 
neighbourhood. In the absence of any $p$-forms, the Einstein-dilaton system 
is found to be asymptotically monotonic and Kasner-like \cite{BK1,AR}.

A convenient tool for describing the qualitative behaviour of the general 
solution near a spacelike singularity is to relate it to billiard motion in 
an auxiliary Lorentzian space, or, after projection on the unit 
hyperboloid, to billiard motion on Lobachevskii space. This billiard 
picture naturally follows from the Hamiltonian approach to cosmological
behaviour (initiated long ago \cite{Misner0} in the context of the Bianchi 
IX models; see \cite{Jantzen} for a recent review), and was first obtained 
in the 4-dimensional case \cite{Chitre,Misnerb} and then extended to higher 
spacetime dimensions with $p$-forms and dilaton 
\cite{Kirillov1993,KiMe,IvKiMe94,IvMe,DH3}. Recent work has improved the 
derivation of the billiard picture by using the Iwasawa decomposition of 
the spatial metric and by highlighting the general mechanism by which all 
the ``off-diagonal'' degrees of freedom ({\it i.e.} all the variables 
except for the scale factors, the dilaton and their conjugate momenta) get 
``asymptotically frozen'' \cite{Damour:2002et}.

The billiard dynamics describing the asymptotic cosmological behaviour of 
$D$-dimensional Einstein gravity coupled to $n$ dilatons $\phi_1 , \ldots , 
\phi_n$ (having minimal kinetic terms $-\Sigma_j \, \vert d \phi_j 
\vert^2$), and to an arbitrary menu of $p$-forms (with kinetic terms $\sim 
- e^{\langle \lambda_p , \phi \rangle} \vert d A_p \vert^2$, where 
$\lambda_p$ is a linear form) consists of a ``billiard ball'' moving along 
a future-directed null geodesic ({\it i.e.} a ``light ray'') of a 
$(D-1+n)$-dimensional 
Lorentzian space, except when it undergoes specular reflections on a set of 
hyperplanar walls (or ``mirrors'') passing through the origin of Lorentzian
space. One can refer to this billiard as being an ``Einstein billiard''
(or a ``cosmological billiard'').

As we shall see below in detail the degrees of freedom of the billiard ball 
correspond, for $D-1$ of them, to the logarithms, $-2 \beta^a$, of the {\it 
scale factors}, {\it i.e.} the diagonal components of the $(D-1) \times 
(D-1)$ spatial metric in an Iwasawa decomposition, and, for the remaining 
ones, to the $n$ dilatons $\phi_j$. The walls come from the asymptotic 
elimination (in the Arnowitt-Deser-Misner  Hamiltonian \cite{ADM})
 of the remaining ``off-diagonal'' degrees
of freedom: off-diagonal components of the metric, $p$-form fields, and 
their conjugate momenta. It is often convenient to project the Lorentzian 
billiard motion (by a linear projection centered on the origin) onto the 
future half of the unit hyperboloid $G_{\mu\nu} \, \beta^{\mu} \, 
\beta^{\nu} = -1$, where $(\beta^{\mu}) \equiv (\beta^a , \phi_j)$ denotes
the position in, and $G_{\mu\nu}$ the metric of $(D-1+n)$-dimensional Lorentzian
space. 
[The signature of $G_{\mu\nu}$ is $-++ \ldots +$.] This projection leads to 
a billiard motion in a domain of hyperbolic space $H_{D-2+n}$, see Fig.~1. 
This domain (the ``billiard table''\footnote{We use the term ``billiard 
table'' to refer to the domain within which the billiard motion is 
confined. Note, however, that this ``table'' is, in general, of dimension 
higher than two, and is either considered in Lorentzian space or, after 
projection, in hyperbolic space.}) is the intersection of a finite number 
of half hyperbolic spaces, say the positive sides of the hyperbolic 
hyperplanar walls defined by intersecting the original Lorentzian 
hyperplanar walls with the unit hyperboloid $G_{\mu\nu} \, \beta^{\mu} \, 
\beta^{\nu} = -1$.

\begin{figure}
\centering \includegraphics{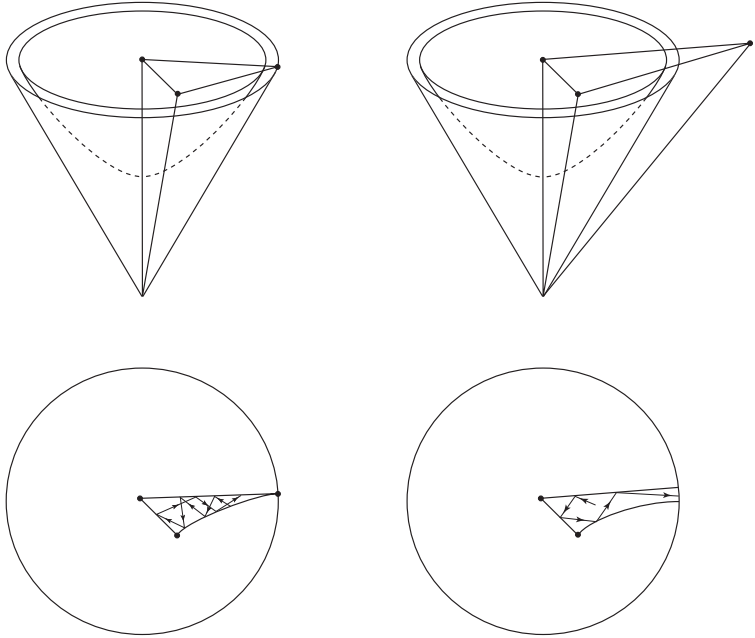}
\caption{Sketch of billiard tables describing the asymptotic cosmological
behaviour of Einstein-matter systems. The upper panels are drawn in the 
$(D-1+n)$--dimensional auxiliary Lorentzian space spanned by
$(\beta^{\mu}) = (\beta^a , \phi)$. [$D=4$ and $n=0$ in the case drawn here.]
The billiard tables are represented as ``wedges'' in [$(2+1)$--dimensional] $\beta$-space,
bounded by hyperplanar walls $w_A (\beta) = 0$ on  which the (unrepresented)
billiard ball undergoes spe\-cular reflections. The upper left panel is a 
(critical) ``chaotic'' billiard table (contained within the $\beta$-space future 
light cone), while the upper right one is a (subcritical) ``non-chaotic'' one 
(extending beyond the light cone). The lower panels represent the corresponding 
billiard tables (and billiard motions) after projection onto hyperbolic space 
[$H_2$ in the case drawn here]. The latter projection is defined in the text by 
central projection onto $\gamma$-space ({\it i.e.} the unit hyperboloid 
$G_{\mu\nu} \, \gamma^{\mu} \, \gamma^{\nu} = -1$, see upper panels), and is 
represented in the lower panels by its image in the Poincar\'e ball [disc]. 
``Chaotic'' billiard tables have finite volume in hyperbolic space, while 
``non-chaotic'' ones have infinite volume.}
\label{Figure1}
\end{figure}

The distinction mentioned above between a chaotic, BKL-like behaviour and a 
monotonic, Kasner-like one then corresponds to the distinction between a 
finite-volume billiard table (in $H_{D-2+n}$) and an infinite-volume one. 
One could further distinguish: (i) the overcritical case ({\it compact} 
billiard table), (ii) the critical case ({\it finite-volume}, but {\it 
non-compact} billiard table) and (iii) the subcritical case ({\it 
infinite-volume} billiard table). Both cases (i) and (ii) lead to 
never-ending, chaotic oscillations of the scale factors (and of the 
dilatons), while case (iii) leads, after a finite number of reflections off 
the walls ({\it i.e.} after a finite number of oscillations of the scale 
factors and of the dilatons), to an asymptotically free motion of 
$\beta^{\mu}$ (null geodesic), correspon\-ding to an asymptotically 
monotonic, Kasner-like (power-law) behaviour of the spacetime metric near 
the singularity, see Fig.~1. In actual physical models, only the critical and 
subcritical cases are found.

Having introduced the concept of (Lorentzian or, after projection, 
hyperbolic) billiard table associated to general classes of Einstein-matter 
systems, we can now describe the connection uncovered in 
\cite{DH1,DH2,DH3,DHJN,DdBHS,DHN2,Damour:2002et} between certain specific 
Einstein-matter systems and Lorentzian Kac-Moody algebras. In the leading 
asymptotic approximation to the behaviour near the cosmological 
singularity, this connection is simply that the (Lorentzian) {\it billiard 
table} describing this behaviour can be identified with the (Lorentzian) 
{\it Weyl chamber} of some corresponding (model-dependent) Lorentzian 
Kac-Moody algebra. We recall that the Weyl chamber of a Lie algebra with 
simple roots\footnote{The roots $\alpha$ are linear forms on a Cartan subalgebra
$\frakh$ that appear as eigenvalues of the adjoint action of $\frakh$
on the rest of the Lie algebra: $ [h, e_{\alpha}] = \alpha(h) e_{\alpha} $
for all $h \in\frakh$ .
In the simple case of the Lie algebra of $SU(2)$, $\frakh = \{ J_z \}$
is one-dimensional and the ``raising'' or ``lowering'' operators
$e_{\alpha} = J_x \pm i J_y$ correspond to the roots $\alpha(J_z) = \pm 1$.}
 $\alpha_i$, $i=1, \ldots , r$ ($r$ being the rank of the
algebra) is the domain of the Cartan subalgebra (parametrized by $\beta \in 
{\mathbb R}^r$) where $\langle \alpha_i , \beta \rangle \geq 0$ for all 
$i$'s. For this connection to be possible many conditions must be met. In 
particular: (i) the billiard table must be a Coxeter polyhedron, {\it i.e.} 
the dihedral angles between adjacent walls must be integer submultiples of 
$\pi$ ({\it i.e.}, of the form $\pi / k$ where $k$ is an integer $\geq 2$), 
and (ii) the billiard table must be a simplex, {\it i.e.} have exactly 
$D-1+n$ faces. It is remarkable that this seemingly very special case of a 
``Kac-Moody billiard'' is found to occur in many physically interesting 
Einstein-matter systems. For instance, pure Einstein gravity in 
$D$-dimensional spacetime corresponds to the Lorentzian Kac-Moody algebra 
$AE_{D-1}$ \cite{DHJN}. The latter algebra is the canonical Lorentzian 
extension \cite{Kac} of the ordinary (finite-dimensional) Lie algebra 
$A_{D-3}$ ($= sl_{D-2}$). Another interesting connection between qualitative PDE behaviour
and Kac-Moody theoretic concepts is that the transition between
``critical'' chaotic, BKL-like behaviour in ``low dimensions'' ($D \leq 
10$) and ``subcritical'' monotonic, Kasner-like behaviour in ``high 
dimensions'' ($D \geq 11$) is found to be in strict correspondence with a 
transition between an {\it hyperbolic} Kac-Moody algebra ($AE_{D-1}$ for $D 
\leq 10$) and a {\it non hyperbolic} one ($AE_{D-1}$ for $D \geq 11$). We 
recall here that V. Kac \cite{Kac} defines a  Kac-Moody algebra to be {\it 
hyperbolic} by the condition that any subdiagram obtained by removing a 
node from its Dynkin diagram be either of finite or affine type. Hyperbolic 
Kac-Moody algebras are necessarily Lorentzian ({\it i.e.} the symmetrizable 
Cartan matrix is of Lorentzian signature), but the reverse is not true in 
general.

Another connection between physically interesting Einstein-matter systems 
and Kac-Moody algebras concerns the low-energy bosonic effective actions 
arising in String and $M$-theory. Bosonic String theory in any spacetime 
dimension $D$ is related to the Lorentzian Kac-Moody algebra $DE_D$ 
\cite{DH3,DdBHS}. The latter algebra is the canonical Lorentzian extension 
of the finite-dimensional algebra $D_{D-2}$. The various Superstring 
theories (in the critical dimension $D=10$) and $M$-theory have been found 
\cite{DH3} to be related either to $E_{10}$ (when there are two 
sypersymmetries in $D=10$, {\it i.e.} for type IIA, type IIB and $M$-theory) or 
to $BE_{10}$ (when there is only one supersymmetry in $D=10$,
{\it i.e.} for type I and the two heterotic theories), see Fig.~2.
See Ref.~\cite{HJ03} for a discussion of the billiards associated to all types
of pure supergravities in $D=4$. See
Ref.~\cite{DdBHS} for the construction of Einstein-matter systems related (in
the above ``billiard'' sense) to the canonical Lorentzian extensions of
{\it all} the finite-dimensional Lie algebras ($A_n$, $B_n$, $C_n$, $D_n$, 
$G_2$, $F_4$, $E_6$, $E_7$ and $E_8$). See also Ref.~\cite{dBS} for the identification
of all the hyperbolic Kac-Moody algebras whose Weyl chambers are Einstein billiards.
\begin{figure}
\centering \includegraphics{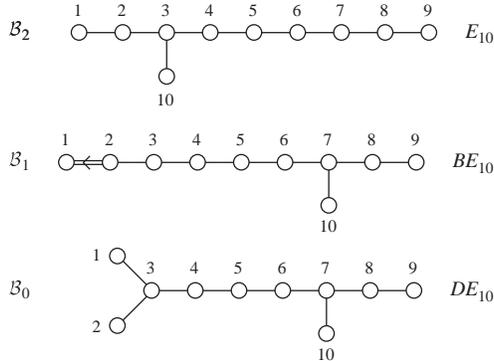}
\caption{Coxeter-Dynkin diagrams encoding the geometry of the billiard tables 
describing the asymptotic cosmological behaviour of three blocks of string 
theories: ${\mathcal B}_2 = \{$$M$-theory, type IIA and type IIB superstring
theories$\}$, ${\mathcal B}_1 = \{$type I and the two heterotic superstring 
theories$\}$, and ${\mathcal B}_0 = \{$closed bosonic string theory in $D=10\}$. 
Each node of the diagrams represents a dominant wall of the corresponding 
cosmological billiard. Each Coxeter diagram of a billiard table corresponds to 
the Dynkin diagram of a correspon\-ding (hyperbolic) Kac-Moody algebra: 
$E_{10}$, $BE_{10}$ and $DE_{10}$, respectively. In other words, the  
cosmological billiard tables can be identified with the Weyl chambers of the 
corresponding Lorentzian Kac-Moody algebras.}
\label{Figure2}
\end{figure}

Ref. \cite{DHN2} has studied in more detail the correspondence between the
specific Einstein-three-form system (including a Chern-Simons term) 
describing the bosonic sector of 11-dimensional supergravity (also known as 
the ``low-energy limit of $M$-theory'') and the hyperbolic Kac-Moody group 
$E_{10}$. %[See also Ref.~\cite{KN1} which uses a different decomposition of the $E_{10}$ algebra.]
Ref. \cite{DHN2} introduced a formal expansion of the field
equations in terms of positive roots, {\it i.e.} combinations $\alpha = 
\Sigma_i \, n^i \, \alpha_i$ of simple roots of $E_{10}$, $\alpha_i$, $i = 
1,\ldots , 10$, where the $n^i$'s are integers $\geq 0$. It is then useful to 
{\it order} this expansion according to the {\it height} of the positive root 
$\alpha = \Sigma_i \, n^i \, 
\alpha_i$, defined as ${\rm ht} (\alpha) = \Sigma_i \, n^i$. The correspondence 
discussed above between the {\it leading} asymptotic evolution near a 
cosmological singularity (described by a billiard) and Weyl chambers of 
Kac-Moody algebras involves only the terms in the field equation whose 
height is ${\rm ht} (\alpha) \leq 1$. By contrast, Ref. \cite{DHN2} could show, 
by explicit calculations, that there exists a way to define, at each 
spatial point $x$, a correspondence between the field variables $g_{\mu\nu} 
(t,x)$, $A_{\mu\nu\lambda} (t,x)$ (and their gradients), and a (finite) 
subset of the parameters defining an element of the (infinite-dimensional) 
coset space $E_{10} / K(E_{10})$ [where $K(E_{10})$ denotes the maximal
compact subgroup of $E_{10}$], such that the (PDE) field equations of 
supergravity get mapped onto the (ODE) equations describing a null geodesic 
in $E_{10} / K(E_{10})$ {\it up to terms of height} 30.

A complementary check of the correspondence between 11-dimensional supergravity
and the $E_{10} / K(E_{10})$ $\sigma$-model has been obtained in \cite{DN04}.
This result was further extended to the correspondence between the
 $E_{10} / K(E_{10})$ $\sigma$-model and, both, {\it massive}  10-dimensional
 IIA supergravity \cite{KN1} and  ten-dimensional IIB supergravity \cite{KN2}.

These tantalizing
results suggest that the infinite-dimensional hyperbolic Kac-Moody group $E_{10}$
may be a ``hidden symmetry'' of supergravity, in the sense of mapping 
solutions onto other solutions (the appearance of $E_{10}$ as possible symmetry group
of supergravity was first hinted at by Julia long ago \cite{Julia,Julia2}).
Note that the conjecture here is that the { \it continuous} group
$E_{10}(\RR)$ be a hidden symmetry group of { \it classical} supergravity.
At the {\it quantum} level, i.e. for M-theory, one expects only a discrete version
of $E_{10}$, say $E_{10}(\ZZ)$, to be a quantum symmetry. See \cite{BGH} 
for recent work trying to
extend the identification of \cite{DHN2} between roots of $E_{10}$ and symmetries
of supergravity/M-theory beyond height 30, and for references about previous suggestions
of a possible role of $E_{10}$. For earlier appearances of the Weyl groups
of the E series in the context of $U$-duality see \cite{LPS,OPR,BFM}. 
A series of recent papers  \cite{West,SWest,SWest2,Englert1,Englert2}
 has also explored the possible role of $E_{11}$
(a non-hyperbolic extension of $E_{10}$) as hidden symmetry of M-theory.
See also \cite{HJP} for another approach to the search for symmetries of
$M$ theory.

 It is also tempting to assume that the 
Kac-Moody groups underlying the other (special) Einstein-matter systems 
discussed above might be hidden (solution-generating) symmetries. For 
instance, in the case of pure Einstein gravity in $D=4$ spacetime, the 
conjecture is that $AE_3$ is such a symmetry of Einstein gravity. [This 
case, and the correspondence between the field variables and the coset ones 
is further discussed in \cite{Damour:2002et}.]

To end this introductory summary, it is important to add a significant 
mathematical caveat. Most of the results mentioned above have been obtained 
by ``physicists' methods'', {\it i.e.} non rigorous arguments. [See, however,
the mathematical results of \cite{XXX} concerning the dynamics of the Bianchi 
type IX homogeneous model.] The only 
exception concerns the ``non chaotic'', monotonic, Kasner-like behaviour of 
the ``subcritical'' systems. The first mathematical proof that a general 
solution of the four-dimensional Einstein-dilaton system has a non chaotic, 
Kasner-like behaviour was obtained by Andersson and Rendall \cite{AR}. By 
using, and slightly extending, the tools of this proof (Theorem 3 of 
\cite{AR}, concerning certain Fuchsian systems), Ref. \cite{DHRW} has then
given mathematical proofs of the Kasner-like behaviour of more general
classes of Einstein-matter systems, notably the $D$-dimensional coupled 
Einstein-dilaton-$p$-form system when the dilaton couplings of the 
$p$-forms belong to the {\it subcritical} domain mentioned above. Ref. 
\cite{DHRW} also gave a proof of the Kasner-like behaviour of pure gravity 
(Ricci tensor $=0$) when the spacetime dimension $D \geq 11$. For other
mathematical studies on ``non-chaotic'' cases see Ref. \cite{IM02}.

An important mathematical challenge is to convert the physicists' arguments 
(summarized in the rest of this review) concerning the ``chaotic billiard'' 
structure of {\it critical} (and {\it over critical}) systems into precise 
mathe\-matical statements. The recent work \cite{DHN2,Damour:2002et}, which 
provides a simplified derivation of the billiard picture and a rather clean 
decomposition of the field variables into ``ODE-described chaotic'' and 
``asymptotically frozen'' pieces, might furnish a useful starting point for 
formulating precise mathematical statements. In this respect, let us note 
two things: (i) a good news is that numerical simulations have fully 
confirmed the BKL chaotic-billiard picture in several models 
\cite{BeGaIs,Berger}, (ii) a bad news is that the physicists' arguments 
might oversimplify the picture by neglecting ``non generic'' spatial points 
where some of the leading walls disappear, because the spatially-dependent 
coefficients measuring the strength of these walls happen to vanish. A 
similar subtlety (vanishing at exceptional spatial points of wall 
coefficients) takes place in the subcritical, non chaotic case (where a 
finite number of collisions on the billiard walls is expected to channel 
the billiard ball in the ``good'', Kasner-like directions). See discussions 
of this phenomenon in \cite{Berger,Rendall:2001nx}.

In the rest of this review, we shall outline the various arguments leading 
to the conclusions summarized above. For a more complete derivation of the 
billiard results see \cite{Damour:2002et}.

\section{Models and Gauge Conditions}

We consider models of the general form
\beq
&&S[{\rm g}_{MN}, \phi, A^{(p)}] = \int d^D x \, \sqrt{- {\rm g}} \;
\Bigg[R ({\rm g}) - \partial_M \phi \partial^M \phi \nonumber \\
&& \hspace{1.5cm} - \frac{1}{2} \sum_p \frac{1}{(p+1)!}
e^{\l_p \phi} F^{(p)}_{M_1 \cdots M_{p+1}} F^{(p)  \, M_1
\cdots M_{p+1}} \Bigg] + \dots \label{keyaction}
\eeq
where units are chosen such that $16 \pi G_N = 1$ (where $G_N$
is Newton's constant) and the spacetime dimension $D \equiv d+1$
is left unspecified. Besides the standard
Einstein-Hilbert term the above Lagrangian contains a dilaton field
$\phi$ and a number of $p$-form fields $A^{(p)}_{M_1 \cdots M_p}$
(for $p\geq 0$). For simplicity, we consider the case where there is only one 
dilaton, {\it i.e.} $n=1$ in the notation of the Introduction. As a convenient 
common
formulation we adopt the Einstein conformal frame and normalize
the kinetic term of the dilaton $\phi$ with weight one with respect to
the Ricci scalar. The Einstein metric ${\rm g}_{MN}$ has Lorentz
signature $(- + \cdots +)$ and is used to lower or raise the
indices; its determinant is denoted by ${\rm g}$. The $p$-form field
strengths $F^{(p)} = dA^{(p)}$ are normalized as
\be
F^{(p)}_{M_1 \cdots M_{p+1}} =
(p+1)  \partial_{[M_1} A^{(p)}_{M_2 \cdots M_{p+1}]} \equiv
\partial_{M_1} A^{(p)}_{M_2 \cdots M_{p+1}} \pm p \hbox{
permutations }.
\ee
The dots in the action (\ref{keyaction}) indicate possible
modifications of the field strength by additional Yang-Mills
or Chapline-Manton-type couplings \cite{pvnetal,CM}, such as
$F_C = dC^{(2)} - C^{(0)} dB^{(2)}$ for two $2$-forms $C^{(2)}$
and $B^{(2)}$ and a $0$-form $C^{(0)}$, as they occur in type
IIB supergravity. Further modifications include Chern-Simons
terms, as in the action for $D=11$ supergravity \cite{CJS}.
The real parameter $\l_p$ measures the strength of the coupling
of $A^{(p)}$ to the dilaton. When $p=0$, we assume that $\l_0\neq 0$
so that there is only one dilaton. This is done mostly for
notational convenience. If there were other dilatons among the
$0$-forms, these should be separated off from the $p$-forms
because they play a distinct r\^ole. They would define additional spacelike
directions in the space of the
(logarithmic) scale factors and  would correspondingly increase the 
dimension of the
relevant hyperbolic billiard.

The metric ${\rm g}_{MN}$, the dilaton field(s)  $\phi$ and the $p$-form 
fields $A^{(p)}_{M_1 \cdots M_p}$ are {\em a priori} arbitrary functions
of both space and time, on which {\it no symmetry conditions are imposed}.
Nevertheless  it will turn out that the evolution equations near the
singularity will be asymptotically the same as those of certain
homogeneous cosmological models. It is important to keep in mind that
this simplification does not follow from imposing extra dimensional 
reduction conditions but emerges as a direct consequence of the general 
dynamics.

Our analysis applies both to past and future singularities, and in
particular to Schwarzschild-type singularities inside black holes.
To follow historical usage, we shall assume for definiteness that
the spacelike singularity lies in the past, at finite distance in
proper time. More specifically, we shall adopt a space-time
slicing such that the singularity ``occurs'' on a constant time
slice ($t= 0$ in proper time). The slicing is built by use of
pseudo-Gaussian coordinates defined by vanishing lapse $N^i = 0$,
with metric 
\be\label{metric1} 
ds^2 = - \big(N(x^0,x^i)
dx^0\big)^2 + g_{ij} (x^0,x^i) dx^i dx^j .
\ee 
For simplicity, we shall work in the following with a {\it coordinate} spatial 
co-frame $dx^i$, as indicated in Eq.~(\ref{metric1}). However, most of what we 
say would go through if we were working in a general {\it time-independent} 
spatial co-frame $\omega^i (x^k) = \omega_j^i (x^k) dx^j$. [The usefulness of 
using a non-coordinate spatial co-frame will show up in the $E_{10}$ case 
discussed below.]
In order to simplify
various formulas later, we shall find it useful to introduce a
rescaled lapse function 
\be\label{metric2} 
\tilde{N}  \equiv N/\sqrt{g} 
\ee 
where $g \equiv {\rm det}\,g_{ij}$. [Note the distinction between spacetime 
quantities ${\rm g}_{MN}$, ${\rm g}$ and space ones $g_{ij}$, $g$. In the 
following, we work only with the spatial metric $g_{ij}$.] We shall see
that a useful gauge, within the Hamiltonian approach, is that
defined by requiring
\be\label{tildeN=rho2}
\tilde{N}= \rho^2,
\ee
where $\rho^2$ is a quadratic combination of the logarithms of the
scale factors and the dilaton(s), which we will define below in
terms of the Iwasawa decomposition of $g_{ij}$.

Throughout this paper, we will reserve the label $t$ for the {\em proper 
time}
\be\label{propertime}
dt = - N dx^0 =  - \tilde{N}\sqrt{g} dx^0,
\ee
whereas the time coordinate associated with the special gauge
Eq.(\ref{tildeN=rho2}) will be designated by $T$, viz.
\be\label{timeT}
dT = - \frac{dt}{\rho^2 \sqrt{g}}.
\ee
Sometimes, it will also be useful to introduce the ``intermediate''
time coordinate $\tau$ that would correspond to the gauge condition
$\tilde{N} =1$.  It is explicitly defined by:
\be\label{timetau}
d\tau = - \frac{dt}{\sqrt{g}} = \rho^2  dT .
\ee
At the singularity the proper time $t$ is assumed to remain finite
and to decrease toward $0^+$. By contrast, the coordinates $T$ and
$\tau$ both increase toward $+\infty$, as ensured by the minus
sign in (\ref{propertime}). Irrespective of the choice of coordinates,
the spatial volume density $g$ is assumed to collapse to zero at each 
spatial
point in this limit.

As for the $p$-form fields, we shall assume, throughout this paper,
a generalized temporal gauge, viz.
\be
\label{temporal}
A^{(p)}_{0 i_2 \cdots i_p} = 0
\ee
where small Latin letter $i,j,...$ denote spatial indices from now on.
This choice leaves the freedom of performing time-independent gauge
transformations, and therefore fixes the gauge only partially.

\section{Asymptotic dynamics in the general case}
\label{general}
\setcounter{equation}{0}
\setcounter{theorem}{0}
\setcounter{lemma}{0}

The main ingredients of the derivation of the billiard picture are:
\begin{itemize}
\item  use of the Arnowitt-Deser-Misner  Hamiltonian formalism \cite{ADM} ,
\item Iwasawa
decomposition of the metric, {\it i.e.} $ g_{ij}\to (\b^a, {\cn^a}_i)$,
\item decomposition of $\b^\m = (\b^a, \phi)$ into radial ($\rho$) and 
angular ($\g^\m$) parts, and \item use of the pseudo-Gaussian gauge
\Ref{tildeN=rho2}, {\it i.e.} of the time coordinate $T$ as the
evolution parameter.
\end{itemize}

More explicitly, with the conventions already described before, we
assume that in some spacetime patch, the metric is given by
(\ref{metric1}) (pseudo-Gaussian gauge), such that the local
volume $g$ collapses at each spatial point as $x^0 \rightarrow
+\infty$, in such a way that the proper time $t$ tends to $0^+$.
We work in the Arnowitt-Deser-Misner  Hamiltonian formalism
\cite{ADM}, i.e. with first order
evolution equations in the phase-space of the system. For
instance, the gravitational degrees of freedom are initially
described by the metric $g_{ij}$ and its conjugate momentum
$\pi^{ij}$. We systematically use the Iwasawa decomposition
of the spatial metric $g_{ij}$\footnote{ If we were working with a 
non-coordinate spatial co-frame $\omega^i = \omega_j^i (x^k) dx^j$, we would use 
the Iwasawa decomposition of the frame components of the spatial metric: 
$d\ell^2 = g_{ij} \omega^i \omega^j$.}. Namely, we write
\be
\label{Iwasawaex}
g_{ij} = \sum_{a=1}^d e^{- 2 \b^a}  {\cn^a}_i  \, {\cn^a}_j
\ee
where $\cn$ is an upper triangular matrix with $1$'s on the diagonal, and 
where we recall that $d \equiv D-1$ denotes the spatial dimension. We shall 
also need the Iwasawa coframe $\{ \theta^a \}$
\be\label{Iwasawa1}
\theta^a = {\cn^a}_i \, dx^i
\ee
as well as the vectorial frame $\{ e_a \}$ dual to the coframe $\{ \theta^a 
\}$:
\be\label{Iwasawa2}
e_a = {\cn^i}_a \frac{\partial}{\partial x^i}
\ee
where the matrix ${\cn^i}_a$ is the inverse of  ${\cn^a}_i$, {\it i.e.},
${\cn^a}_i {\cn^i}_b = \delta^a_b$.  It is again an upper
triangular matrix with 1's on the diagonal.

The Iwasawa decomposition allows us to replace the $d(d+1)/2$ variables
$g_{ij}$ by the $ d + d(d-1)/2$ variables $(\b^a, {\cn^a}_i)$.
Note that $(\b^a, {\cn^a}_i)$ are {\it ultralocal} functions of
$g_{ij}$, that is they depend, at each spacetime point, only on
the value of $g_{ij}$ at that point, not on its derivatives. This
would not have been the case if we had used a ``Kasner frame''
({\it i.e.} a frame diagonalizing $\pi_{ij}$ with respect to $g_{ij}$)
instead
of an Iwasawa one. The transformation $g \to (\b, \cn)$ then
defines a corresponding transformation of the conjugate momenta,
as we will explain below. We then augment the definition of the
$\b$'s by adding the dilaton field, {\it i.e.} $ \b^\m  \equiv (\b^a,
\phi)$. The $d+n$ dimensional space of the $\beta^{\mu}$'s (where, 
as we said, we consider only one dilaton, {\it i.e.} $n=1$) comes equipped 
with a canonical Lorentzian metric $G_{\mu\nu}$ defined by
\be\label{DeWittbis}
d \sigma^2 \equiv G_{\m \n} d\b^{\m} d\b^{\n} \equiv \sum_{a=1}^d  (d 
\b^a)^2 -  (\sum_{a=1}^d  d\b^a)^2 + d\phi^2 . 
\ee
This (flat) Lorentzian metric in the auxiliary $\beta$-space (of signature 
$-+ \ldots +$) plays an essential role in our study. Note that in models without 
dilaton (such as pure gravity in spacetime dimension $d+1$) one has a 
$d$-dimensional Lorentzian space, with $(\beta^{\mu}) \equiv (\beta^a)$ and 
metric $G_{\mu\nu} \, d\beta^{\mu} \, d\beta^{\nu} = \sum_a (d\beta^a)^2 - 
(\sum_a d\beta^a)^2$.

We then decompose $\b^{\m}$ in hyperbolic polar coordinates
$(\rho, \g^{\m})$, {\it i.e}
\begin{equation}
\label{hyperbolic} \beta^\mu = \rho \gamma^\mu,
\end{equation}
where $\gamma^\mu$ are coordinates on the future sheet of the unit
hyperboloid\footnote{Indeed, one finds that, near a spacelike singularity 
$\beta^{\mu}$ tends to future timelike infinity.}, which are constrained by
\begin{equation}
G_{\mu\nu} \, \gamma^{\mu} \, \gamma^{\nu} \equiv \gamma^\mu \gamma_\mu = -1
\end{equation}
and $\rho$ is the timelike variable
\be\label{rhosquare}
\rho^2 \equiv -G_{\mu\nu} \, \beta^{\mu} \, \beta^{\nu} \equiv - \b^\mu 
\b_\mu > 0 .
\ee
This decomposition naturally introduces the unit hyperboloid 
(``$\g$-space''), see Fig.~1, which is a realization of the $m$-dimensional 
hyperbolic 
(Lobachev\-skii) space $H_m$, with $m=d-1+n\;$ if there are $n\geq0$ 
dilatons.

In terms of the ``polar'' coordinates $\rho$ and $\gamma^\mu$, the metric 
in $\b$-space becomes
\begin{equation}
d\s^2 = - d \rho^2 + \rho^2 d\S^2
\end{equation}
where $d\S^2$ is the metric on the $\g$-space $H_m$.

Note that $\rho$ is also an ultralocal function
of the configuration variables $(g_{ij}, \phi)$. We assume that
the hyperbolic coordinate $\rho$ can be used everywhere in a given
region of space near the singularity as a well-defined (real)
quantity which tends to $+ \infty$ as we approach the singularity.
We then define the slicing of spacetime by imposing the gauge
condition \Ref{tildeN=rho2}. The time coordinate corresponding to
this gauge is called $T$ as above (see \Ref{timeT}). Our aim is to
study the asymptotic behaviour of all the dynamical variables
$\b(T), \cn(T),.....$ as $T \to + \infty$ (recall that this limit
also corresponds to $ t \to 0$, $\sqrt{g} \to 0$, $\rho \to +
\infty$, with $\b^\m$ going to infinity inside the future light
cone). Of course, we must also ascertain the self-consistency of
this limit, which we shall refer to as the ``BKL limit''.

\subsection{Arnowitt-Deser-Misner Hamiltonian action}

To focus on the features relevant to the billiard picture, we
assume here that there are no Chern-Simons terms or couplings of
the exterior form gauge fields through a modification of the
curvatures $F^{(p)}$, which are thus taken to be Abelian, $F^{(p)}
= d A^{(p)}$. See \cite{Damour:2002et} for a proof that these
interaction terms do not change the analysis. The Arnowitt-Deser-Misner Hamiltonian
action in any pseudo-Gaussian gauge, and in the temporal gauge
\Ref{temporal}, reads \cite{ADM}
\beq
&& S\left[ g_{ij}, \pi^{ij}, \phi, \pi_\phi, A^{(p)}_{j_1 \cdots j_p},
\pi_{(p)}^{j_1 \cdots j_p}\right] = \nonumber \\
&& \hspace{1cm}
\int dx^0 \int d^d x \left( \pi^{ij} \dot{g_{ij}}
+ \pi_\phi \dot{\phi} + \frac{1}{p!}\sum_p \pi_{(p)}^{j_1 \cdots j_p}
\dot{A}^{(p)}_{j_1 \cdots j_p} - H \right)
\label{GaussAction}
\eeq
where the Hamiltonian density $H$ is
\beq\label{Ham}
H &\equiv&  \tilde{N} \ch \\
\ch &=& \ck + \cm \\
\ck &=& \pi^{ij}\pi_{ij} - \frac{1}{d-1} \pi^i_{\;i} \pi^j_{\;j}
+ \frac{1}{4} \pi_\phi^2 + \nonumber \\
&& \hspace{2cm} + \sum_p \frac{e^{- \lambda_p \phi}} {2 \, p!}
\, \pi_{(p)}^{j_1 \cdots j_p}
\pi_{(p) \, j_1 \cdots j_p} \\
\cm &=& - g R + g g^{ij} \partial_i \phi \partial_j \phi + \sum_p
\frac{e^{ \lambda_p \phi}}{2 \; (p+1)!} \, g \, F^{(p)}_{j_1
\cdots j_{p+1}} F^{(p) \, j_1 \cdots j_{p+1}}
\eeq
where $R$ is the spatial curvature scalar. The dynamical
equations of motion are obtained by varying the above action
with respect to the spatial metric components, the dilaton, the spatial
$p$-form components and their conjugate momenta. In addition,
there are constraints on the dynamical variables,
\beq
\ch &\approx& 0  \; \; \; \; \hbox{(``Hamiltonian constraint")}, \\
\ch_i &\approx& 0  \; \; \; \; \hbox{(``momentum constraint")}, \\
\varphi_{(p)}^{j_1 \cdots j_{p-1}} &\approx& 0 \; \; \; \;
\hbox{(``Gauss law" for each $p$-form) } \label{Gauss}
\eeq
with
\beq
\ch_i &=& -2 {\pi^j}_{i|j} + \pi_\phi
\partial_i \phi + \sum_p \frac1{p!} \
\pi_{(p)}^{j_1 \cdots j_p} F^{(p)}_{i j_1 \cdots j_{p}} \\
\varphi_{(p)}^{j_1 \cdots j_{p-1}} &=&
{\pi_{(p)}^{j_1 \cdots j_{p-1} j_p}}_{\vert j_p}
\eeq
where the subscript $|j$ stands for the spatially covariant derivative.

Let us now see how the Hamiltonian action gets transformed when one
performs, at each spatial point, the Iwasawa decomposition
\Ref{Iwasawaex} of the spatial metric. The kinetic terms of the metric and 
of the dilaton in the Lagrangian (\ref{keyaction}) are given by the quadratic 
form
\be\label{dsigma2}
\sum_{a=1}^d (d \beta^a)^2 - \left(\sum_{a=1}^d d \beta^a\right)^2  + 
d\phi^2 + \frac{1}{2} \sum_{a<b} e^{2(\beta^b - \beta^a)}
\left( {d\cn^a}_i \, {\cn^i}_b \right)^2
\ee
where we recall that ${\cn^i}_a$  denotes, as in  \Ref{Iwasawa2},
the inverse of the triangular matrix ${\cn^a}_i$ appearing in the
Iwasawa decomposition  \Ref{Iwasawaex} of the spatial metric $g_{ij}$.
This change of variables corresponds to a point canonical transformation,
which can be extended to the momenta in the standard way via
\be\label{cantra}
\p^{ij}\dot{g}_{ij} \equiv \sum_a \pi_a \dot{\b}^a
+ \sum_{a} {P^i}_a \dot{{\cal N}^a}_{i} .
\ee
Note that the momenta
\be\label{Nmomenta}
{P^i}_a = \frac{\partial\cal L}{\partial \dot{{\cal N}^a}_i}
= \sum_{b} e^{2(\beta^b - \beta^a)} {\dot\cn^a}_{\;\;j} {\cn^j}_b {\cn^i}_b,
\ee
conjugate to the non-constant off-diagonal Iwasawa components
${\cn^a}_i$, are only defined for $a<i$; hence the second sum in
(\ref{cantra}) receives only contributions from $a<i$.

We next split the Hamiltonian  density\footnote{We use the term
``Hamiltonian density'' to denote both $H$ and $\ch$. Note that
$H$ is a usual spatial density (of weight 1, {\it i.e.} the same weight 
as $\sqrt{g}$), while $\ch \equiv \sqrt{g} H/N$ is  a density 
of weight 2 (like $ g = (\sqrt{g})^2$). Note also that 
$\p^{ij}$ is of weight 1, while $ \tilde{N} \equiv N/\sqrt{g}$
is of weight $-1$.} $\ch$  (\ref{Ham}) in two parts, one
denoted by ${\cal H}_0$, which is the kinetic term for the local
scale factors $\beta^\mu$ (including dilatons) already encountered
in section 3, and a ``potential density'' (of weight 2) denoted by
$\cv$, which contains everything else.
Our analysis below will show why it makes sense to group the kinetic
terms of both the off-diagonal metric components and the $p$-forms
with the usual potential terms, {\it i.e.} the term $\cal M$ in (\ref{Ham}).
[Remembering that, in a Kaluza-Klein reduction, the off-diagonal
components of the metric in one dimension higher become a
one-form, it is not surprising that it might be useful to group
together the off-diagonal components and the $p$-forms.]
Thus, we write
\be\label{HplusV}
\ch =  {\cal H}_0 + \cv
\label{calH}
\ee
with the kinetic term of the $\b$ variables
\be\label{eq3.23}
{\cal H}_0 = \frac{1}{4} G^{\mu\nu} \pi_\mu \pi_\nu
\ee
where $G^{\mu\nu}$ denotes the inverse of the metric $G_{\mu\nu}$ of 
Eq.~(\ref{DeWittbis}). In other words the right hand side of Eq.~(\ref{eq3.23}) 
is defined by
\begin{equation}\label{Gmunuup}
G^{\mu \nu} \pi_\mu \pi_\nu \equiv
\sum_{a=1}^d \pi_a^2 - \frac{1}{d-1} \left(\sum_{a=1}^d
\pi_a\right)^2 + \pi_\phi^2
\end{equation}
where $ \pi_\mu \equiv (\pi_a, \pi_\phi)$ are the
momenta conjugate to $\beta^a$ and $\phi$, respectively, {\it i.e.}
\be
\pi_\m = 2 \tilde{N}^{-1} G_{\m \n} \dot{\beta}^\n =
2 G_{\m \n} \frac { d {\beta}^\n}{d\tau} .
\ee
The total (weight 2) potential density,
\be
\cv = \cv_S + \cv_G + \sum_p \cv_{p}  + \cv_\phi ,
\ee
is naturally split into a ``centrifugal'' part linked to the kinetic
energy of the off-diagonal components (the index ``$S$'' referring
to ``symmetry'', as discussed below)
\be
\label{centrifugal}
\cv_S = \frac{1}{2}
\sum_{a<b} e^{-2(\beta^b - \beta^a)}
\left( {P^j}_b {{\cal N}^a}_j\right)^2,
\ee
a ``gravitational'' (or ``curvature'') potential
\be
\label{gravitational}
\cv_G =  - g R ,
\ee
and a term from the $p$-forms,
\be
 \cv_{(p)} = \cv_{(p)}^{el} + \cv_{(p)}^{magn}
\ee
which is a sum of an ``electric'' and a ``magnetic'' contribution
\beq
\cv_{(p)}^{el} &=&  \frac{e^{- \lambda_p \phi}} {2 \, p!} \,
\pi_{(p)}^{j_1 \cdots j_p}
\pi_{(p) \, j_1 \cdots j_p} \\
\cv_{(p)}^{magn} &=&  \frac{e^{ \lambda_p \phi}}{2 \; (p+1)!} \,
g \, F^{(p)}_{j_1 \cdots j_{p+1}} F^{(p) \, j_1 \cdots j_{p+1}} .
\eeq
Finally, there is a  contribution to the potential linked to the
spatial gradients of the dilaton:
\be
\cv_\phi  = g g^{ij} \partial_i \phi \partial_j \phi.
\ee
We will analyze below in detail these contributions to the potential.

\subsection{Appearance of sharp walls in the BKL limit}

In the decomposition of the Hamiltonian given above, we have split
off the kinetic terms of the  scale factors $\b^a$ and of the dilaton
$\b^{d+1} \equiv \phi$ from the other variables, and assigned
the off-diagonal metric components and the $p$-form fields to
various potentials, each of which is a complicated function of
$\b^\m, {\cn^a}_i, {P^i}_a,  A^{(p)}_{j_1 \cdots j_{p}},
\pi_{(p)}^{j_1 \cdots j_p}$ and of some of their spatial gradients.
The reason why this separation is useful is that, as we are going to show,
in the BKL limit, and in the special Iwasawa  decomposition which
we have adopted, the asymptotic dynamics is governed by the scale
factors $\b^\m$, whereas all other variables ``freeze''.
Thus, in the asymptotic limit, we have schematically
\be
\cv\Big( \b^\m, {\cn^a}_i, {P^i}_a,  A^{(p)}_{j_1 \cdots j_{p}},
\pi_{(p)}^{j_1 \cdots j_p},\dots\Big) \longrightarrow \cv_{\infty}(\g^\m)
\ee
where $\cv_{\infty}(\g^\m)$ stands for a sum of certain
``sharp wall potentials'' which depend only on the angular hyperbolic
coordinates $\g^\m \equiv \b^\m/ \r$. As a consequence, the asymptotic
dynamics can be described as a ``billiard'' in the hyperbolic space
of the $\g^\m$'s, whose walls (or ``cushions'') are determined by the
energy of the fields that are asymptotically frozen.

This reduction of the complicated potential to a much simpler
``effective potential'' $\cv_{\infty}(\g^\m)$ follows essentially
from the exponential dependence of $\cv$ on the diagonal Iwasawa
variables $\b^\m$, from its independence from the conjugate
momenta of the $\b$'s, and from the fact that the radial magnitude
$\rho$ of the $\b$'s becomes infinitely large in the BKL limit.

To see the essence of this reduction, with a minimum of technical
complications, let us consider a potential density (of weight 2) 
of the general form
\be\label{V1}
\cv(\beta, Q, P) = \sum_A c_A(Q, P) \exp\big(- 2 w_A (\beta) \big)
\ee
where $(Q,P)$ denote the remaining phase space variables (that is,
other than $(\b, \p_\b)$). Here $w_A (\beta) = w_{A \m} \b^\m$
are certain linear forms which depend only on the (extended) scale factors,
and whose precise form will be derived in the following section. Similarly
we shall discuss below the explicit form of the pre-factors $c_A$, which
will be some complicated polynomial functions of the remaining
fields, {\it i.e.} the off-diagonal components of the metric, the $p$-form
fields and their respective conjugate momenta, and of some of their
spatial gradients.  The fact that the $w_A (\beta)$ depend
{\em linearly} on the scale factors $\beta^\m$ is an important property
of the models under consideration. A second non-trivial fact is that,
{\em for the leading contributions}, the pre-factors are always
non-negative, {\it i.e.} $c_A^{\rm leading}\geq 0$. Since the values of the
fields for which $c_A =0$ constitute a set of measure zero, we will
usually make the ``genericity assumption'' $c_A >0$ for the leading terms
in the potential $\cv$\footnote{As mentioned in the Introduction 
understanding the effects of the possible failure of
this assumption is one of the subtle issues in establishing
a rigorous proof of the BKL picture.}. The third fact following from
the detailed analysis of the walls that we shall exploit is that all
the leading walls are {\it timelike}, {\it i.e.} their normal vectors
(in the Minkowski $\b$-space) are spacelike.

When parametrizing $\beta^{\mu}$ in terms of $\rho$ and $\gamma^{\mu}$, or 
equivalently
\be
\label{lambda}
\lambda  \equiv \ln \rho \equiv \ft12 \ln \big( -G_{\m \n} \b^\m \b^\n\big)
\ee
(with conjugate momentum $\pi_{\lambda}$), and $\gamma^{\mu}$, the part of 
the Hamiltonian des\-cribing the kinetic energy of the $\beta$'s, $H_0 = 
\tilde N {\mathcal H}_0$, takes the form
\begin{equation}
H_0 = \frac{\tilde{N}}{4\rho^2} \left[- \pi_\lambda^2 + \pi_{\bf
\gamma}^2\right].
\label{Hamiltonianrho}
\end{equation}
Choosing the gauge \Ref{tildeN=rho2} to simplify
the kinetic terms $H_0$ we end up with an Hamiltonian of the form
\beq\label{V2}
H\big(\lambda, \pi_{\lambda}, \g, \pi_{\g},Q,P\big) &=& \tilde{N} \ch \\
&& \!\!\!\!\!\!\!\!\!\!\!\!\!\!\!\!\!\!\!\!\!\!\!\!
= \frac{1}{4} \left[- \pi_\lambda^2 + \pi_{\bf \gamma}^2\right] +
\rho^2 \sum_A c_A(Q,P) \exp\big(-\rho w_A (\gamma) \big) \nonumber
\eeq
where $\pi_{\bf \gamma}^2$ is the kinetic energy of a particle
moving on $H_m$. In \Ref{V2} and below we shall regard $\lambda$
 as a primary dynamical variable
(so that $\r \equiv e^\l$).

The essential point is now that, in the BKL limit, $\l \to + \infty$ {\it 
i.e.} $\r \to + \infty$, each term $\rho^2 \exp\big(- 2 \rho w_A (\gamma) 
\big)$ becomes a {\it sharp wall potential}, {\it i.e.} a function of  $w_A 
(\gamma)$ which is zero when $w_A (\gamma) >0$, and  $+\infty$ when $w_A 
(\gamma) < 0$. To formalize this behaviour we define the sharp wall 
$\Theta$-function\footnote{One should more properly write 
$\Theta_\infty(x)$, but since this is the only step function encountered in 
this article, we use the simpler notation $\Theta(x)$.} as
\be
\Theta (x) := \left\{ \begin{array}{ll}
                      0  & \mbox{if $x<0$} \\
                      +\infty & \mbox{if $x>0$}
                      \end{array}
                      \right.
\ee 
A basic formal property of this $\Theta$-function is its
invariance under multiplication by a positive quantity. With the
above assumption checked below that all the relevant prefactors
$c_A(Q,P)$ are {\it positive} near each leading wall, we can
formally write 
\be 
\lim_{\rho\rightarrow\infty} \Big[ c_A(Q,P)
\rho^2 \exp\big(-\rho w_A (\gamma) \Big] = c_A(Q,P)\Theta\big(-2
w_A (\gamma) \big) \equiv \Theta\big(- 2 w_A (\gamma) \big). 
\ee
Of course, $\Theta (-2 w_A (\gamma)) = \Theta (- w_A (\gamma))$,
but we shall keep the extra factor of $2$ to recall that the
arguments of the exponentials, from which the $\Theta$-functions
originate, come with a well-defined normalization. Therefore, the
limiting Hamiltonian density reads 
\be \label{V3}
H_{\infty}(\lambda, \p_{\lambda}, \g, \pi_{\g},Q,P) =
 \frac{1}{4} \left[- \pi_\lambda^2 + \pi_{\bf
\gamma}^2\right] +
\sum_{A'} \Theta\big(-2 w_{A'} (\gamma) \big),
\ee
where $A'$ runs over the {\it dominant walls}.  The set of dominant walls
is defined as the {\it minimal set} of wall forms which suffice to define
the billiard table, {\it i.e.} such that the restricted set of inequalities 
$  \lbrace w_{A'} (\gamma) \geq 0 \rbrace$ imply the full set $   \lbrace 
w_{A} (\gamma) \geq 0 \rbrace$.  The concept of dominant wall will be 
illustrated below. [Note that the concept of ``dominant'' wall is a 
refinement of the distinction, which  will also enter our discussion, 
between a leading wall and a subleading  one.]

The crucial point is that the limiting Hamiltonian \Ref{V3} no longer 
depends on $\lambda, Q$ and $P$. Therefore the Hamiltonian equations of 
motion for $\lambda, Q$ and $P$ tell us that  the corresponding conjugate 
momenta, {\it i.e.} $\p_{\lambda}, P $ and $Q$, respectively, all become
{\it constants of the motion} in the limit $\lambda \to +\infty$.
The total Hamiltonian density $H_{\infty}$ is also a constant
of the motion (which must be set to zero). The variable $\lambda$
evolves according to $d \lambda/ dT = - \ft12 \p_{\lambda}$. Hence,
in the limit, $\lambda$ is a linear function of $T$. The only non-trivial
dynamics resides in the evolution of $(\g, \p_{\g})$ which
reduces to the sum of a free (non relativistic)
kinetic term  $\p_{\g}^2/4$ and a sum of sharp wall potentials,
such that the resulting motion of the $\gamma$'s
indeed constitutes a billiard, with geodesic motion on the unit
hyperboloid $H_m$ interrupted by reflections on the walls defined by
$w_A(\gamma)=0$. These walls are hyperplanes (in the sense of
hyperbolic geometry) because they are geometrically given by
the intersection of the unit hyperboloid $\b^\mu \b_\mu =-1$
with the usual Minkowskian hyperplanes $w_A(\b)=0$.

See Appendix A of \cite{Damour:2002et} for the study of a toy model which 
shows in more detail how the asymptotic constancy of the ``off diagonal'' 
phase-space variables $(Q,P)$ arises.

Geodesic motion in a billiard in hyperbolic space has been much
studied. It is known that this motion is chaotic or non-chaotic
depending on whether the billiard has finite or infinite volume
\cite{Ma69,HM79,Z84,Esk}. In the finite volume case, the generic
evolution exhibits an infinite number of collisions with the walls
with strong chaotic features (``oscillating behavior").

By contrast, if the billiard has infinite volume, the evolution is
non-chaotic.  For a generic evolution, there are only finitely
many collisions with the walls.  The system generically settles
after a finite time in a Kasner-like motion that lasts all the way
to the singularity, see Fig.~1.

The above derivation relied on the use of hyperbolic polar coordinates
$(\rho,\gamma)$. This use is technically useful in that it represents the
walls as being located at an asymptotically  fixed position in hyperbolic
space, namely $w_A(\gamma)=0$. However, once one has derived the final 
result \Ref{V3}, one can reexpress it in terms of the original variables 
$\b^\m$, which run over a linear (Minkowski) space. Owing to the linearity 
of $w_A (\beta) = \rho w_A (\gamma)$ in this Minkowskian picture, the
asymptotic motion takes place in a ``polywedge'', bounded by the 
hyperplanes $w_A(\b)=0$. The billiard motion then consists of free motions 
of  $\b^\m$ on straight lightlike lines within this polywedge, which are 
interrupted by specular reflections off the walls. [See formula 
\Ref{collision} below for the explicit effect of these reflections on the 
components of the velocity vector of the $\b$-particle.] Indeed, when going 
back to $\b$-space ({\it i.e.} before taking the BKL limit), the dynamics 
of the scale factors at each point of space is given by the Hamiltonian
\be
\label{V4}
H(\b^\m, \p_{\m}) =
\tilde{N} \ch = {\tilde{N}}   \left[\ft14 G^{\mu \nu} \pi_\mu \pi_\nu  +
\sum_A c_A   \exp\big(- 2 w_A (\b)  \big)    \right] .
\ee
The $\b$-space dynamics simplifies in the gauge $\tilde{N} =1$, 
corresponding
to the time coordinate $\tau$. In the BKL limit, the Hamiltonian \Ref{V4}
takes the limiting form (in the gauge $ \tilde{N} = 1$)
\be\label{V5}
H_{\infty}(\b^\m, \p_{\m}) = \ft14 G^{\mu \nu} \pi_\mu \pi_\nu  +
\sum_{A'}    \Theta \big(- 2 w_{A'} (\b)  \big)
\ee
where the sum is again only over the dominant walls.
When taking ``equal time'' slices of this polywedge ({\it e.g.}, slices on
which $\Sigma_i \b^i$ is constant), it is clear that with increasing
time ({\it i.e} increasing $\Sigma_i \b^i $, or increasing $\tau$ or 
$\rho$)
the walls recede  from the observer. The $\b$-space  picture
is useful for simplifying the mathematical representation of the dynamics
of the scale factors which takes place in a linear space. However, it is
inconvenient both for proving that the exponential walls of \Ref{V4}
do reduce, in the large $\Sigma_i \b^i $ limit to sharp walls, and
for dealing with the dynamics of the other phase-space variables
$(Q,P)$, whose appearance in the coefficients  $ c_A $ has been suppressed
in \Ref{V4} above. Let us only mention that, in order to prove,
in this picture, the freezing of the  phase space variables $(Q,P)$
one must consider in detail the accumulation of the  ``redshifts''
of the energy-momentum $\p_\m$ of the $\b$-particle when it undergoes
reflections on the receding walls, and the effect of the resulting
decrease of the magnitudes of the components of $\p_\m$ on the
evolution of $(Q,P)$.
By contrast, the $\g$-space picture that we used above allows for a more
streamlined treatment of the effect of the limit $\r \to + \infty$
on the sharpening of the walls and on the freezing of $(Q,P)$.

In summary, the dynamics simplifies enormously in the asymptotic
limit. It becomes ultralocal in that it reduces to a continuous
superposition of  evolution systems (depending only on a time
parameter) for the scale factors and the dilatons, at each spatial
point, with asymptotic freezing of the off-diagonal and $p$-form
variables. This ultralocal description of the dynamics is valid
only asymptotically. It would make no sense to speak of a billiard
motion prior to this limit, because one cannot replace the
exponentials by $\Theta$-functions. Prior to this limit, the
evolution system for the scale factors involves the coefficients
in front of the exponential terms, and the evolution of these
coefficients depends on various spatial gradients of the other
degrees of freedom. However, one may contemplate setting up an
expansion in which the sharp wall model is replaced by a model
with exponential (``Toda-like") potentials, and where the
evolution of the quantities entering the coefficients of these
``Toda walls'' is treated as a next to leading effect. See
\cite{DHN2} and below for the definition of the first steps of such an
expansion scheme for maximal supergravity in eleven dimensions.

\subsection{Constraints}
\label{constraints}

We have just seen that in the BKL limit, the evolution equations
become ordinary differential equations with respect to time.
Although the spatial points are decoupled in the evolution
equations, they are, however, still coupled via the constraints.  These
constraints just restrict the initial data and need only be
imposed at one time, since they are preserved by the dynamical
equations of motion. Indeed, one easily finds that, {\it in the BKL limit},
\be
\dot{\cal H } = 0
\ee
since $[{\cal H}(x), {\cal H}(x')] = 0$ in the
ultralocal limit.  This corresponds simply to the fact that the
collisions preserve the lightlike character of the velocity
vector. Furthermore, the gauge constraints (\ref{Gauss}) are also
preserved in time since the Hamiltonian constraint is gauge-invariant.
{\it In the BKL limit}, the momentum constraint fulfills
\be
\dot{{\cal H}}_k(x) = \partial_k {\cal H} \approx 0 .
\ee
It is important that the restrictions on the initial data do not bring
dangerous constraints on the coefficients of the walls in the sense
that these may all take non-zero values. For instance, it is well
known that it is consistent with the Gauss law to take non-vanishing
electric and magnetic energy densities; thus the coefficients of
the electric and magnetic walls are indeed generically non-vanishing.
In fact, the constraints are essentially conditions on the
spatial gradients of the variables entering the wall coefficients,
not on these variables themselves.  In some non-generic contexts,
however, the constraints could force some of the wall coefficients
to be zero; the corresponding walls would thus be absent. [{\it E.g.},
for vacuum gravity in four dimensions, the momentum constraints
for some Bianchi homogeneous models force some symmetry wall
coefficients to vanish.  But this is peculiar to the homogeneous
case.]

It is easy to see that the number of arbitrary physical functions
involved in the solution of the asymptotic  BKL equations of motion is the
same as in the general solution of the complete Einstein-matter
equations.  Indeed, the number of constraints on the initial data
and the residual gauge freedom are the same in both cases. Further
discussion of the constraints in the BKL context may be found in
\cite{AR,DHRW}.

\subsection{Consistency of BKL behaviour in spite of the increase
of spatial gradients}

The essential assumption in the BKL analysis is the asymptotic dominance of time 
derivatives with respect
to space derivatives near a spacelike singularity.
This assumption has been mathematically justified, in a rigorous manner,
in the  cases where  the billiard is of infinite volume, {\it i.e.} in the
(simple) cases where the asymptotic behaviour is not chaotic,
but is  monotonically Kasner-like \cite{AR,DHRW}.

On the other hand, one might a priori worry that this assumption
is self-contradictory in those cases where the billiard is of
finite volume, when the asymptotic behaviour is chaotic, with an
infinite number of oscillations.  Indeed, it has been pointed out
\cite{turb1,turb2} that the independence of the billiard evolution
at each spatial point will have the effect of {\it infinitely
increasing} the spatial gradients of various quantities, notably
of the local values of the Kasner exponents $p_\m(x)$. This
increase of spatial gradients towards the singularity has been
described as a kind of turbulent behaviour of the gravitational
field, in which energy is pumped into shorter and shorter length
scales  \cite{turb1,turb2}, and, if it were too violent, it would
certainly work against the validity of the BKL assumption of
asymptotic dominance of time derivatives. For instance, in our
analysis of gravitational walls in the following section, we will
encounter subleading walls, whose prefactors depend on spatial
gradients of the logarithmic scale factors $\b$.

To address the question of consistency of the BKL assumption
we need to know how fast the spatial gradients of $\b$, and of similar
quantities entering  the prefactors, grow near the singularity.
We refer to \cite{Damour:2002et} for a discussion of this issue. The result is
that the chaotic
character of the billiard indeed implies an unlimited growth of
the spatial gradients of $\b$, but that this growth is only of {\it
polynomial} order in $\r$ 
\be
\partial_i \b = {\cal O}(1)  \r^2 , 
\ee
where the coefficient ${\cal O}(1)$ is a chaotically oscillating quantity. This 
polynomial growth of $\partial_i \b$ (and of its second-order
spatial derivatives) entails a polynomial growth of the prefactors
of the sub-dominant walls. Because it is
polynomial (in $\r$), this growth  is, however, negligible compared
to the {\it exponential} (in $\r$) behaviour of the various
potential terms. It does not jeopardize our reasoning based on
keeping track of the various exponential behaviours. As we will
see the potentially dangerously growing terms that we have
controlled here appear only in subdominant walls. The reasoning of
the Appendix of \cite{Damour:2002et} shows that the prefactors of the dominant 
walls are
self-consistently predicted to evolve very little near the
singularity.

We conclude that the unlimited growth of some of the spatial gradients
does not affect the consistency of the BKL analysis done here.
This does not mean, however, that it will be easy to promote our
analysis to a rigorous mathematical proof. The main obstacle to such
a proof appears to be the existence of exceptional points, where a 
prefactor of a dominant wall happens to vanish, or points
where a subdominant wall happens to be comparable to a dominant one.
Though the set of such exceptional  points is (generically) of measure
zero, their density might increase near the singularity
because of the increasing spatial gradients. This situation
might be compared to the KAM (Kolmogorov-Arnold-Moser) one, where the
``bad'' tori have a small measure, but are interspersed densely among
the ``good'' ones.

\section{Walls}
\setcounter{equation}{0}
\setcounter{theorem}{0}
\setcounter{lemma}{0}

The decomposition (\ref{HplusV}) of the Hamiltonian gives rise to
different types of walls, which we now discuss in turn.
Specifically, we will derive explicit formulas for the linear
forms $w_A(\b)$ and the field dependence of the pre-factors
$c_A$ entering the various  potentials.

\subsection{Symmetry walls}
\label{Geometryoff} We start by analyzing the effects of the
off-diagonal metric components which will give rise to the
so-called ``symmetry walls''. As they originate from the
gravitational action they are always present. The relevant
contributions to the potential is the centrifugal potential
\Ref{centrifugal}. When comparing  \Ref{centrifugal} to the
general form \Ref{V1} analyzed above, we see that firstly the
summation index $A$ must be interpreted as a double index $(a,b)$,
with the restriction $ a<b$, secondly that the corresponding
prefactor is $ c_{ab} =   ({P^j}_a {{\cal N}^b}_j )^2$ is
automatically non-negative (in accordance with our genericity
assumptions, we shall assume $ c_{ab} >  0$). The centrifugal wall
forms read: \be w_{(ab)}^S(\b) \equiv w_{(ab)\mu}^S \beta^\mu
\equiv \beta^b - \beta^a  \; \; (a<b). \label{symwall} \ee We
refer to these wall forms as the ``symmetry walls'' for the
following reason.  When applying the general collision law
\Ref{collision} derived below to the case of the collision on the
wall \Ref{symwall} one easily finds that its effect on the
components of the velocity vector $v^\m$ is simply to permute the
components $v^a$ and $v^b$, while leaving unchanged the other
components $\m \ne a,b$.

The hyperplanes  $w_{(ab)}^S(\beta) = 0$ ({\it i.e.} the symmetry walls)
are timelike since 
\be 
G^{\mu \nu} w_{(ab) \mu}^S w_{(ab) \nu}^S
=+ 2 \label{normofgrad} .
\ee 
This ensures that the symmetry walls
intersect the hyperboloid $G_{\m \n} \b^\m \b^\n = -1$, $\sum_a
\beta^a \geq 0$. The symmetry billiard (in $\b$-space)  is defined
to be the region of  Minkowski space determined by  the
inequalities 
\be
w_{(ab)}^S(\beta) \geq 0, 
\label{fullsymmetry}
\ee 
with $\sum_a \beta^a\geq 0$ ({\it i.e.} by the region of $\b$-space
where the $\Theta$ functions are zero). Its projection on the
hyperbolic space $H_m$ is defined by the inequalities $
w_{(ab)}^S(\g) \geq 0$.

The explicit expressions above of the symmetry wall forms also allow us to
illustrate the
notion of a ``dominant wall'' defined above.
Indeed, the  $d(d-1)/2$ inequalities
\Ref{fullsymmetry} already follow from the following  {\it minimal} set of
$d-1$ inequalities
\be
\b^2 -\b^1 \geq 0, \; \b^3 - \b^2 \geq 0,
\; \cdots , \b^d - \b^{d-1} \geq 0 .
\label{symmetry}
\ee
More precisely, each linear form which must be positive
in \Ref{fullsymmetry} can be written as a linear combination,
with positive (in fact, integer) coefficients of the linear
forms entering the subset \Ref{symmetry}. For instance,
$ \b^3 - \b^1 = (\b^3 - \b^2) + (\b^2 -\b^1) $, etc. 
As discussed in \cite{Damour:2002et} this result can be reinterpreted by 
identifying the
dominant linear forms entering \Ref{symmetry} with the simple
roots of $ SL(n,\RR)$. 

\subsection{Curvature (gravitational) walls}

Next we analyze the gravitational potential, which requires a
computation of curvature. To that end, one must explicitly express
the spatial curvature in terms of the scale factors and the
off-diagonal variables ${\cn^a}_i$. Again, the calculation is most
easily done in the Iwasawa frame (\ref{Iwasawa1}). We use the short-hand 
notation $A_a \equiv e^{-\b^a}$ for the (Iwasawa) scale factors. Let $C^a_{\; \; 
bc}(x)$ be the structure functions of the Iwasawa basis $\{\theta^a\}$, 
viz.
\be
d \theta^a = - \frac{1}{2}C^a_{\; \; bc} \theta^b\wedge \theta^c
\ee
where $d$ is the spatial exterior differential. The structure
functions obviously depend only on the off-diagonal components
${{\cal N}^a}_i$, but not on the scale factors.
Using the Cartan formulas for the connection one-form
$\o^a_{\; \; b}$,
\beq
&& d\theta^a + \sum_b \o^a_{\; \; b} \wedge \theta^b = 0 \\
&& d \gamma_{ab} = \o_{ab} + \o_{ba}
\eeq
where $ \o_{ab}\equiv \gamma_{ac} {\o^c}_b$, and
\be
\gamma_{ab} = \delta_{ab} A_a^2\equiv \exp(-2\b^a) \delta_{ab}
\ee
is the metric in the frame $\{\theta^a\}$, one finds
\beq
\o^c_{\; \; d} &=& \sum_b \frac{1}{2} \left( C^b_{\; \; cd}
\frac{A_b^2}{A_c^2} + C^d_{\; \; cb} \frac{A_d^2}{A_c^2}
 - C^c_{\; \; db}  \frac{A_c^2}{A_d^2} \right) \theta^b \nonumber \\
&& + \sum_b \frac{1}{2 A^2_c} \Big[ \delta_{cd} (A^2_c)_{,b}
+ \d_{cb} (A^2_c)_{,d} - \d_{db} (A^2_d)_{,c} \Big] \theta^b .
\eeq
In the last bracket above, the commas denote the frame
derivatives   $\partial_a \equiv {\cn^i}_a \partial_i$.

The Riemann tensor $R^c_{\; \; def}$, the Ricci tensor $R_{de}$
and the scalar curvature $R$ are obtained through
\beq
\Omega^a_{\; \; b} &=& d \o^a_{\; \; b}
+ \sum_c \o^a_{\; \; c} \wedge \o^c_{\; \; b} \\
&=& \frac{1}{2} \sum_{e,f} R^a_{\; \; bef} \theta^e \wedge \theta^f
\eeq
where $\Omega^a_{\; \; b}$ is the curvature $2$-form and
\be
R_{ab} =  \sum_c R^c_{\; \; acb}, \; \; \;
R = \sum_a \frac{1}{A^2_a} R_{aa}.
\ee
Direct, but somewhat cumbersome, computations yield
\be
R = - \frac{1}{4} \sum_{a,b,c} \frac{A_a^2}{A^2_b \, A^2_c}
(C^a_{\; \;bc})^2 + \sum_a \frac{1}{A_a^2} F_a\big(\partial^2 \beta,
\partial \beta, \partial C, C\big)
\label{formulaforR}
\ee
where $F_a$ is some complicated function of its arguments
whose explicit form will not be needed here.  The only property
of $F_a$ that will be of importance is that it is a polynomial
of degree two in the derivatives $\partial \beta$ and of degree
one in $\partial^2 \beta$.  Thus, the exponential dependence
on the $\beta$'s which determines the asymptotic behaviour
in the BKL limit, occurs only through the $A^2_a$-terms written
explicitly in (\ref{formulaforR}).

In \Ref{formulaforR} one obviously has $b\neq c$ because the
structure functions $C^a_{\; \; bc}$ are antisymmetric in the
pair $[bc]$. In addition to this restriction, we can assume,
without loss of generality, that $a \not=b,c$ in the first
sum on the right-hand side of (\ref{formulaforR}). Indeed, the
terms with either $a=b$ or $a=c$ can be absorbed into a
redefinition of $F_a$. We can thus write the gravitational
potential density (of weight 2) as
\be\label{VsubG}
\cv_G \equiv -g R =
\frac{1}{4} {\sum_{a,b,c}}' e^{-2\a_{abc}(\beta)}
(C^a_{\; \; bc})^2 - \sum_a e^{-2 \m_a(\beta)} F_a
\ee
where the prime on $\sum$ indicates that the sum is to be performed
only over unequal indices, i.e. $a\neq b, b\neq c , c\neq a$, and
where the linear forms $\a_{abc}(\beta)$ and $\m_a(\beta)$
are given by
\be
\a_{abc}(\beta) = 2 \beta^a + \sum_{e\neq a,b,c} \beta^e \quad\quad
(a\neq b \, , \; b\neq c \, , \; c\neq a)
\ee
and
\be
\m_a(\beta) = \sum_{c \not= a} \beta^c
\ee
respectively. Note that $ \a_{abc}$ is symmetric under the exchange
of $b$ with $c$, but that the index $a$ plays a special role.

Comparing the result \Ref{VsubG} to the general form    \Ref{V1}
we see that there are, a priori, two types of gravitational walls:
the $\a$-type and the $\m$-type. It is shown in \cite{Damour:2002et} that 
the $\alpha$-walls dominate the $\mu$-walls and, more precisely that $\mu_a 
(\beta) \geq 0$ within the entire future light cone of the $\beta$'s. [The 
proof uses the fact that each linear form $\mu_a (\beta)$ is lightlike.]

{}From these considerations we deduce the additional constraints
\be
\a_{abc}(\b) \geq 0 \; \; \;  \; (D>3)
\ee
besides the symmetry inequalities (\ref{symmetry}). The hyperplanes
$\a_{abc}(\b) = 0$ are called the ``curvature" or ``gravitational"
walls. Like the symmetry walls, they are timelike since
\be
G^{\mu\nu} (\a_{abc})_\mu (\a_{abc})_\nu = +2 .
\ee
The restriction $D>3$ is due to the fact that in $D=3$ spacetime
dimensions, the gravitational walls $\a_{abc}(\b) = 0$ are absent,
simply because one cannot find three distinct spatial indices. 
This is, of course, in agreement with expectations, because gravity
in three spacetime dimensions has no propagating degrees of
freedom (gravitational waves).

\subsection{$p$-form walls}

While none of the wall forms considered so far involved the dilatons,
the electric and magnetic ones do as we shall now show. To make the
notation less cumbersome we will omit the super-(or sub-)script $(p)$
on the $p$-form fields in this subsection.

\subsubsection{Electric walls}
The electric potential density can be written as
\be
\cv^{el}_{(p)} =
\frac{1}{2 \, p!} \sum_{a_1, a_2, \cdots, a_p} e^{-2 e_{a_1  \cdots
a_p}(\b)} (\ce^{a_1  \cdots a_p})^2
\label{Velec}
\ee
where $\ce^{a_1  \cdots a_p}$ are the components of the electric field
$\pi^{i_1 \cdots i_p}$ in the basis $\{\theta^a\}$
\be
\ce^{a_1 \cdots a_p} \equiv {\cn^{a_1}}_{j_1}
{\cn^{a_2}}_{j_2} \cdots {\cn^{a_p}}_{j_p} \pi^{j_1 \cdots j_p}
\ee
(recall our summation conventions for spatial cordinate indices)
and where $e_{a_1  \cdots a_p}(\b)$ are the electric wall
forms\footnote{A sign error in the last term $\propto \l_p$ crept in
the corresponding equation of \cite{Damour:2002et}.}
\be
e_{a_1 \cdots a_p}(\b) = \b^{a_1} + \cdots + \b^{a_p} +
\frac{\l_p}{2} \phi  .
\ee
Here the indices $a_j$'s are all distinct because
$\ce^{a_1 \cdots a_p}$ is completely antisymmetric. The variables
$\ce^{a_1\cdots a_p}$ do not depend on the $\beta^\mu$. It is thus rather
easy to take the BKL limit. The exponentials in (\ref{Velec}) are
multiplied by positive factors which generically are different
from zero. Thus, in the BKL limit, $\cv^{el}_{(p)}$ becomes
\be
\cv^{el}_{(p)} \simeq \sum_{a_1< a_2< \cdots< a_p} \Theta[-2
e_{a_1  \cdots a_p}(\b)] .
\ee
The transformation from the variables $\big({\cn^a}_i,
{P^i}_a, A_{j_1 \cdots j_p}, \pi^{j_1 \cdots j_p}\big)$ to
the variables $\big({\cn^a}_i,{{\cal P}^i}_a, {\cal A}_{a_1 \cdots a_p},
\ce^{a_1  \cdots a_p}\big)$ is a point canonical
transformation whose explicit form is obtained from
\be
\sum_{a} {P^i}_a \dot{\cn^a}_{i} + \sum_p \frac1{p!}
\pi^{j_1 \cdots j_p} \dot{A}_{j_1 \cdots j_p} =
\sum_{a}{{\cal P}^i}_a \dot{\cn^a}_i +
\sum_p\sum_{a_1,\dots,a_p}\frac1{p!}
\ce^{a_1 \cdots a_p} \dot{{\cal A}}_{a_1 \cdots a_p} .
\ee
The new momenta ${{\cal P}^i}_a$   conjugate to ${\cn^a}_{i}$ differ from 
the old ones  ${P^i}_a$  by terms involving $\ce$,  $\cn$
and $\ca$ since the components ${\cal A}_{a_1 \cdots a_p}$ of
the $p$-forms in the basis $\{\theta^a\}$ depend on the $\cn$'s,
$$
{\cal A}_{a_1 \cdots a_p} =
{\cn^{j_1}}_{a_1} \cdots {\cn^{j_p}}_{a_p} A_{j_1\cdots j_{p}}.
$$
However, it is easy to see that these extra terms do not affect
the symmetry walls in the BKL limit.

\subsubsection{Magnetic walls}
The magnetic potential is dealt with similarly.  Expressing it
in the $\{\theta^a\}$-frame, one obtains
\be
\cv^{magn}_{(p)} =
\frac{1}{2 \, (p+1)!} \sum_{a_1, a_2, \cdots, a_{p+1}} e^{-2
m_{a_{1}  \cdots a_{p+1}}(\b)} (\cf_{a_1  \cdots a_{p+1}})^2
\label{Vmagn}
\ee
where $\cf_{a_1  \cdots a_{p+1}}$ are the components
of the magnetic field $F_{m_1 \cdots m_{p+1}}$ in the basis
$\{\theta^a\}$,
\be
\cf_{a_1 \cdots a_{p+1}} =
{\cn^{j_1}}_{a_1} \cdots  {\cn^{j_{p+1}}}_{a_{p+1}} F_{j_1 \cdots j_{p+1}} .
\ee
The $m_{a_{1} \cdots a_{p+1}}(\b)$ are the magnetic linear
forms\footnote{A sign error in the last term $\propto \l_p$ crept in
the corresponding equation of \cite{Damour:2002et}.}
\be
m_{a_{1} \cdots a_{p+1}}(\b) = \sum_{b \notin \{a_1,a_2,\cdots a_{p+1}\}}
\b^b - \frac{\l_p}{2} \phi
\ee
where again all $a_j$'s are distinct. One sometimes rewrites
$m_{a_{1} \cdots a_{p+1}}(\b)$ as $\tilde{m}_{a_{p+2} \cdots a_d}$,
where $\{a_{p+2}, a_{p+3}, \cdots, a_d \}$ is the set complementary
to \break $\{a_1,a_2,\cdots a_{p+1} \}$; {\it e.g.},
\be
\tilde{m}_{1 \, 2 \, \cdots \, d-p-1} =
\b^1 + \cdots + \b^{d-p-1} -\frac{\l_p}{2} \phi
= m_{d-p \, \cdots \, d} .
\ee
Of course, the components of the exterior derivative $\cf$ of $\ca$ in the 
non-holonomic frame $\{\theta^a\}$ involves the structure coefficients,
{\it i.e.} $\cf_{a_1  \cdots a_{p+1}} = \partial_{[a_1} {\cal A}_{a_2 
\cdots a_{p+1}]} + C {\cal A}\hbox{-terms}$ where $\partial_a \equiv 
{\cn^i}_a \partial_i$ is the frame derivative.

Again, the BKL limit is quite simple and yields (assuming generic
magnetic fields)
\be
\cv^{magn}_{(p)} \simeq \sum_{a_1< \cdots < a_{d-p-1}}
 \Theta[-2 \tilde{m}_{a_1 \cdots a_{d-p-1}}(\b)].
\ee
Just as the off-diagonal variables, the electric and magnetic
fields freeze in the BKL limit since the Hamiltonian no longer
depends on the $p$-form variables. These drop out because one can
rescale the coefficient of any $\Theta$-function to be one (when
it is not zero), thereby absorbing the dependence on the $p$-form
variables.

The scale factors are therefore constrained by the further 
``billiard'' conditions
\be
e_{a_1  \cdots a_p}(\b) \geq 0, \; \; \;
\tilde{m}_{a_1 \cdots  a_{d-p-1}}(\b) \geq 0.
\ee
The hyperplanes $e_{a_1  \cdots a_p}(\b) = 0$ and
$\tilde{m}_{a_1 \cdots  a_{d-p-1}}(\b) =0$ are called ``electric"
and ``magnetic" walls, respectively. Both walls are timelike
because their gradients are spacelike, with squared norm
\be
\frac{p(d-p-1)}{d-1} + \Big(\frac{\lambda_p}{2}\Big)^2 > 0 .
\label{elecnorm}
\ee
(For $D=11$ supergravity, we have $d=10, p=3$ and $\lambda_p = 0$
and thus the norm is equal to $+2$.) This equality explicitly shows
the invariance of the norms of the $p$-form walls under
electric-magnetic duality.

\subsection{Subleading walls}

The fact that the leading walls originating from the centrifugal,
gravitational and $p$-form or are all timelike, is an important
ingredient of the overall  BKL picture. We refer to \cite{Damour:2002et}
for a discussion of the other contributions to the Hamiltonian (including 
Chern-Simons and Yang-Mills couplings) and for a proof that they contribute 
only subleading walls.

\section{Einstein (or cosmological) billiards}
\setcounter{equation}{0}
\setcounter{theorem}{0}
\setcounter{lemma}{0}

Let us summarize the findings above. The dynamics in the vicinity of
a spacelike singularity is governed by the scale factors,
while the other variables (off-diagonal metric components,
$p$-form fields) tend to become mere ``spectators'' which get 
asymptotically frozen. This simple result is most easily derived in terms
of the hyperbolic polar coordinates ($\r,\g$), and in the
gauge \Ref{tildeN=rho2}. In this picture, the essential dynamics
is carried by the angular variables $\g$ which move on a fixed
billiard table, with cushions defined by the dominant walls
$w_{A'}(\g)$. One can refer to this billiard as being an ``Einstein billiard''
(or a ``cosmological billiard'').

It is often geometrically more illuminating
to ``unproject'' the billiard motion in the full Minkowski space
of the extended scale factors $\b^\m$.  In that picture, the asymptotic
evolution of the scale factors at each spatial point reduces to a
zigzag of null straight lines with respect to the metric $G_{\mu \nu} d\b^\m
d\b^\n$. The straight segments of this motion are interrupted by collisions
against the sharp walls
\be\label{cushions}
w_A(\b) \equiv w_{A \m} \b^\m = 0
\ee
defined by the symmetry, gravitational and $p$-form potentials,
respectively. As we showed all these walls are {\it timelike},
i.e. they have spacelike gradients:
\be
G^{\m\n} w_{A \m}  w_{A \n} > 0 .
\ee
Indeed, the gradients of the symmetry and gravitational wall forms
have squared norm  equal to $+2$, independently of the dimension $d$.
By contrast, the norms of the electric and magnetic gradients, which
are likewise positive, depend on the model. As we saw, there also
exist subdominant walls, which can be neglected as they are
located ``behind'' the dominant walls.

In the $\b$-space picture, the free motion before a
collision is described by a null straight line. The effect
of a collision on a particular wall $w_A(\b)$ is easily obtained
by solving \Ref{V4}, or \Ref{V5}, with only one term in the sum.
This dynamics is exactly integrable: it suffices to decompose the
motion of the $\b$-particle into two (linear) components: $(i)$ the
component parallel to the (timelike) wall hyperplane, and $(ii)$ the
orthogonal component. One easily finds that the parallel motion
is left unperturbed by the presence of the wall, while the orthogonal 
motion suffers a (one-dimensional) reflection, with a change of the sign of 
the out\-going orthogonal velocity with respect to the ingoing one. The net
effect of the collision on a certain wall  $w(\b)$ then is to change the 
ingoing velocity vector $v^\m = d\beta^{\mu} / d\tau$ entering the ingoing free 
motion 
into an outgoing velocity vector $v'^\m$ given (in any 
linear frame) by the usual formula for a geometric reflection in the  
hyperplane $w(\b) =0$:
\begin{equation}
v'^{\mu} =  v^{\mu} - 2 \ \frac{(w \cdot v) \, w^{\mu}}{(w \cdot w)}  \, .
\label{collision}
\end{equation}
Here, all scalar products, and index raisings, are done with the $\b$-space
metric $G_{\m\n}$. Note that the collision law \Ref{collision} leaves
invariant the (Minkowski) length of the vector $ v^{\mu}$. Because the
dominant walls are timelike, the geometric reflections that the velocities
undergo during a collision, are elements of the orthochronous Lorentz
group. Each reflection preserves the norm and the time-orientation;
hence, the velocity vector remains null and future-oriented.

{}From this perspective, we can also better understand the relevance of
walls which are {\em not} timelike. {\em Lightlike} walls (like some
of the subleading gravitational walls) can never cause reflections
because in order to hit them the billiard ball would have to move
at superluminal speeds in violation of the Hamiltonian constraint.
The effect of {\em spacelike} walls (like the cosmological constant
wall) is again different: they are either irrelevant (if they are
``behind the motion''), or otherwise they reverse the time-orientation
inducing a motion towards increasing spatial volume (``bounce'').

The {\it hyperbolic billiard} is obtained from the $\b$-space
picture by a radial projection onto the unit hyperboloid of the 
piecewise straight motion in the polywedge defined by the walls. 
The straight motion thereby becomes a
geodesic motion on hyperbolic space. The ``cushions'' of the
hyperbolic billiard table are the intersections of the hyperplanes
\Ref{cushions} with the unit hyperboloid, such that the billiard
motion is constrained to be in the region defined as the intersection
of the half-spaces $w_A(\b) \geq 0$ with the unit hyperboloid.
As we already emphasized, not all walls are relevant since some
of the inequalities $w_A(\b) \geq 0$ are implied by others
\cite{DH3}. Only the dominant wall forms, in terms of which all
the other wall forms can be expressed as linear combinations with
non-negative coefficients, are relevant for determining the
billiard. Usually, these are the minimal symmetry walls and some of the
$p$-form walls. The billiard region, as a subset of hyperbolic
space, is in general {\it non-compact} because the cushions meet at
infinity ({\it i.e.} at a cusp); in terms of the original scale factor
variables $\b$, this means that the corresponding hyperplanes
intersect on the lightcone. It is important that, even when the
billiard is non-compact, the hyperbolic region can have finite volume.

Given the action \Ref{keyaction} with definite spacetime
dimension, menu of fields and dilaton couplings, one can determine
the relevant wall forms and compute the billiards.  For generic
initial conditions, we have the following results, as to which of
the models (\ref{keyaction}) exhibit oscillatory behaviour (finite
volume billiard) or Kasner-like behaviour (infinite volume billiard)
\begin{itemize}
\item Pure gravity billiards have finite volume for spacetime dimension $D
\leq 10$ and infinite volume for spacetime dimension $D \geq 11$
\cite{DHS}. This can be understood in terms of the underlying
Kac-Moody algebra $AE_d$ \cite{DHJN}: as shown there, the system is chaotic
precisely if the underlying indefinite Kac-Moody algebra is hyperbolic.
\item The billiard of gravity coupled to a dilaton always has infinite
volume, hence exhibits Kasner-like behavior  \cite{BK1,AR,DHRW}.
\item If gravity is coupled to $p$-forms (with $p \not = 0$ and $p <D-2$)
{\em without a dilaton} the corresponding billiard has a finite
volume \cite{DH2}. The most prominent example in this class is
$D=11$ supergravity, whereas vacuum gravity in $11$ dimensions
is Kasner-like. The $3$-form is crucial for closing the billiard.
Similarly, the Einstein-Maxwell system in four (in fact any number of)
dimensions has a finite-volume billiard (see \cite{Jantzen0,Leblanc,Weaver}
for a discussion of four-dimensional homogeneous models with Maxwell
fields exhibiting oscillatory behaviour).
\item The volume of the mixed Einstein-dilaton-$p$-form system depends
on the dilaton couplings. For a given spacetime dimension $D$ and
a given menu of $p$-forms there exists a subcritical open domain
$\cal D$ in the space of the dilaton couplings, i.e. an open
neighbourhood of the origin $\lambda_p = 0$ such that:
$(i)$ when the dilaton couplings $\lambda_p$ belong to $\cal D$
the general behaviour is Kasner-like, but $(ii)$ when the $\lambda_p$
do not belong to $\cal D$ the behaviour is oscillatory \cite{DH1,DHRW}.
For all the superstring models, the dilaton couplings do not belong to
the subcritical domain and the billiard has finite volume.  Note, however, 
that the superstring dilaton couplings are precisely ``critical'', {\it 
i.e.} on the borderline between the subcritical and the overcritical 
domain.
\end{itemize}

As a note of caution let us point out that some indicators of chaos
must be used with care in general relativity, because of reparametrization
invariance, and in particular redefinitions of the time coordinate;
see \cite{Cor,Imp} for a discussion of the original Bianchi IX model. In 
this respect, we refer to \cite{Damour:2002et} for a discussion of the link 
between the various time coordinates used in the analysis above: $t$, 
$\tau$ and $T$.

The hyperbolic billiard description of the (3+1)-dimensional homogeneous
Bianchi IX system was first worked out by Chitre \cite{Chitre}
and Misner \cite{Misnerb}. It was subsequently generalized to
inhomogeneous metrics in \cite{Kirillov1993,IvKiMe94}.
The extension to higher dimensions with perfect fluid sources was
considered in \cite{KiMe}, without symmetry walls. Exterior $p$-form
sources were investigated in \cite{IvMe,Ivashchuk:1999rm} for special
classes of metric and $p$-form configurations. The uniform approach (based 
on the Iwasawa decomposition) summarized above comes from 
\cite{Damour:2002et}.

\section{Kac-Moody theoretic formulation: the $E_{10}$ case}
\label{KM0} \setcounter{equation}{0} \setcounter{theorem}{0}
\setcounter{lemma}{0}

Although the billiard description holds for all systems governed
by the action (\ref{keyaction}), the billiard in general has no notable
regularity property. In particular, the dihedral angles between
the faces, which can depend on the (continuous) dilaton couplings,
need not be integer submultiples of $\pi$. In some instances,
however, the billiard can be identified with the fundamental Weyl
chamber of a symmetrizable Kac-Moody (or KM) algebra of indefinite 
type\footnote{Throughout this section, we will use the abbreviations
``KM'' for ``Kac-Moody'', and ``CSA'' for Cartan subalgebra.}, with 
Lorentzian signature metric \cite{DH3,DHJN,DdBHS}. Such billiards are
called ``Kac-Moody billiards''. More specifically, in \cite{DH3},
superstring models were considered and the rank $10$ KM algebras
$E_{10}$ and $BE_{10}$ were shown to emerge, in line with earlier
conjectures made in \cite{Julia,Julia2}\footnote{Note that the Weyl
groups of the $E$-family have been discussed in a similar vein in
the context of $U$-duality \cite{LPS,OPR,BFM}.}. This result was further
extended to pure gravity in any number of spacetime dimensions, for
which the relevant KM algebra is $AE_d$, and it was understood that
chaos (finite volume of the billiard) is equivalent to hyperbolicity
of the underlying Kac-Moody algebra \cite{DHJN}. For pure gravity
in $D=4$ the relevant algebra is the hyperbolic algebra $AE_3$
first investigated in \cite{FF}. Further examples of emergence of
Lorentzian Kac-Moody algebras, based on the models of \cite{BGM,CJLP3},
are given in \cite{DdBHS}. See also \cite{dBS} and \cite{HJ03}.

The main feature of the gravitational billiards that can be associated
with KM algebras is that there exists a group theoretical interpretation
of the billiard motion: the asymptotic BKL dynamics is equivalent (in a 
sense
to be made precise below), at each spatial point, to the asymptotic
dynamics of a one-dimensional nonlinear $\sigma$-model based on a certain
infinite dimensional coset space $G/K$, where the KM group $G$ and its 
maximal compact
subgroup $K$ depend on the specific model. As we have seen, the walls
that determine the billiards are the {\it dominant walls}. For
KM billiards, they correspond to the {\it simple roots} of the KM algebra.
As we discuss below, some of the subdominant walls also have an algebraic 
interpretation in
terms of higher-height positive roots. This enables
one to go beyond the BKL limit and to see the beginnings of a possible
identification of the dynamics of the scale factors {\em and} of all
the remaining variables with that of a non-linear $\sigma$-model
defined on the  cosets of the Kac-Moody group divided its maximal
compact subgroup \cite{DHN2,Damour:2002et}.

The KM theoretic reformulation not only enables us to give a
unified group theoretical derivation of the different types of walls
discussed in the preceding section, but also shows that
{\em the $\beta$-space of logarithmic scale factors, in which
the billard motion takes place, can be identified
with the Cartan subalgebra of the underlying indefinite
Kac-Moody algebra.} The various types of walls can thus be
understood directly as arising from the large field limit
of the corresponding $\s$-models. It is the presence of gravity, which
comes with a metric in scale-factor space of Lorentzian signature,
which forces us to consider {\it infinite dimensional} groups
if we want to recover all the walls found in our previous analysis,
and this is the main reason we need the theory of KM algebras.
For finite dimensional Lie algebras we obtain only a subset of
the walls: one of the cushions of the associated billiard is missing, 
and one always ends up with a monotonic
Kasner-type behavior in the limit $t\rightarrow 0^+$. The absence of 
chaotic oscillations
for models based on finite dimensional Lie groups is consistent 
with the classical
integrability of these models. While they remain formally integrable
for infinite dimensional KM groups, one can understand the
chaotic behavior as resulting from the projection of a motion in
an infinite dimensional space onto a finite dimensional subspace.

For concreteness, we shall only consider one specific example here: the
relation between the cosmological evolution of $D=11$ supergravity and a 
null geodesic on $E_{10} / K(E_{10})$ \cite{DHN2}. See Refs.~\cite{KN1,KN2}
for the extension of this correspondence to the two supergravities in $D=10$,
and Ref.~\cite{DN04} for a complementary check of the correspondence in the
case of $D=11$ supergravity. We refer to
\cite{Damour:2002et} for a more general discussion of the link between KM
billiards and KM coset models (including a discussion of the $AE_3$ case 
relevant for pure Einstein gravity in $3+1$ dimensions). Note also that the 
relevance of non-linear $\s$-models for uncovering the symmetries of 
$M$-theory has also been discussed from a different, spacetime-covariant 
point of view in \cite{West,SWest,SWest2}, but there it is $E_{11}$ rather 
than $\E$ that has been proposed as a fundamental symmetry.

The action defining the bosonic part of $D=11$ supergravity reads
\begin{eqnarray}
\label{sugra}
S &= &\int d^{11} x \Bigl[ \sqrt{-{\rm G}} \, R({\rm G}) - \frac{\sqrt{-{\rm 
G}}}{48} \, {\mathcal F}_{\alpha\beta\gamma\delta} \, {\mathcal 
F}^{\alpha\beta\gamma\delta} \nonumber \\
&&+ \frac{1}{(12)^4} \, \varepsilon^{\alpha_1 \ldots \alpha_{11}} \, {\mathcal 
F}_{\alpha_1 \ldots \alpha_4} \, {\mathcal F}_{\alpha_5 \ldots \alpha_8} \, 
{\mathcal A}_{\alpha_9 \alpha_{10} \alpha_{11}} \Bigl]
\end{eqnarray}
where the spacetime indices $\alpha , \beta , \ldots = 0,1, \ldots , 10$, where 
$\varepsilon^{01 \ldots 10} = +1$, and where the four-form ${\mathcal F}$ is the 
exterior derivative of ${\mathcal A}$: ${\mathcal F} = d {\mathcal A}$. Note the 
presence of the Chern-Simons term ${\mathcal F} \wedge {\mathcal F} \wedge 
{\mathcal A}$ in the action (\ref{sugra}). Introducing a zero-shift slicing 
($N^i=0$) of the eleven-dimensional
spacetime, and a {\em time-independent} spatial zehnbein
$\theta^a(x) \equiv {E^a}_i(x) dx^i$, the metric and four form
${{\cal F}} = d\cA$ become
\begin{eqnarray}\label{Gauge}
  &&  ds^2 = {\rm G}_{\alpha\beta} \, dx^{\alpha} \, dx^{\beta} = - N^2 
(d{x^0})^2 + G_{ab} \theta^a \theta^b \\ \!\!\!\!\!\!
{{\cal F}} &=& \frac1{3!}{{\cal F}}_{0abc}\,  dx^0 
\!\wedge\!\theta^a\!\wedge\!
\theta^b\!\wedge\! \theta^c
+ \frac1{4!}{{\cal F}}_{abcd} \, \theta^a\!\wedge\!\theta^b\!\wedge\!
\theta^c\!\wedge\!\theta^d . \nn
\end{eqnarray}
We choose the time coordinate $x^0$ so that the lapse $N=\sqrt{G}$,
with $G:= \det G_{ab}$ (note that $x^0$ is not the proper time\footnote{In this 
section we denote the proper time by $T$ to keep the variable $t$ for denoting 
the parameter of the one-dimensional $\sigma$-model introduced below.}
$T = \int N dx^0$;
rather, $x^0\rightarrow\infty$ as $T \rightarrow 0$). In this frame
the complete evolution equations of $D=11$ supergravity read
\begin{eqnarray}\label{EOM}
\partial_0 \big( G^{ac} \partial_0 G_{cb} \big)  &=&
\ft16 G {{\cal F}}^{a\beta\gamma\delta} {{\cal F}}_{b\beta\gamma\delta} -
\ft1{72} G {{\cal F}}^{\alpha\beta\gamma\delta} {{\cal 
F}}_{\alpha\beta\gamma\delta} \delta^a_b \non
   &&        - 2 G {R^a}_b (\Gamma,C) \non
\partial_0 \big( G{{\cal F}}^{0abc}\big) &=&
\ft1{144} \varepsilon^{abc a_1 a_2 a_3 b_1 b_2 b_3 b_4}
          {{\cal F}}_{0a_1 a_2 a_3} {{\cal F}}_{b_1 b_2 b_3 b_4} \non
  &&  \!\! \!\! \!\! \!\! \!\! \!\! \!\! \!\!
 + \ft32 G {{\cal F}}^{de[ab} {C^{c]}}_{de} - G {C^e}_{de} {{\cal F}}^{dabc}
     - \partial_d \big( G{{\cal F}}^{dabc} \big) \non
\partial_0 {{\cal F}}_{abcd} &=& 6 {{\cal F}}_{0e[ab} {C^e}_{cd]} + 4 
\partial_{[a} {{\cal F}}_{0bcd]}
\end{eqnarray}
where $a,b \in \{1,\dots,10\}$ and $\alpha,\beta \in \{0,1,\dots,10\}$,
and $R_{ab}(\Gamma,C)$ denotes the spatial Ricci tensor; the (frame)
connection components are given by
$
2 G_{ad} {\Gamma^d}_{bc} = C_{abc} + C_{bca} - C_{cab} +
          \partial_b G_{ca} + \partial_c G_{ab} - \partial_a G_{bc}
$
with ${C^a}_{bc} \equiv G^{ad} C_{dbc}$ being the structure coefficients of
the zehnbein $d\theta^a = \frac12 {C^a}_{bc} \theta^b \!\wedge\! \theta^c$.
[Note the change in sign convention here compared to above.] The frame 
derivative is $\partial_a \equiv {E^i}_a (x) \partial_i$ (with
$ {E^a}_i {E^i}_b = \delta^a_b$). To determine the solution at any {\it 
given}
spatial point $x$ requires knowledge of an infinite tower
of spatial gradients: one should thus
augment \Ref{EOM} by evolution equations for
$\partial_a G_{bc}, \partial_a {{\cal F}}_{0bcd}, \partial_a {{\cal 
F}}_{bcde}$, 
etc.,
which in turn would involve higher and higher spatial gradients.

The geodesic Lagrangian on $\E/K(\E)$ is defined by generalizing
the standard Lagrangian on a finite dimensional coset space $G/K$,
where $K$ is a maximal compact subgroup of the Lie group $G$.
All the elements entering the construction of ${{\cal L}}$ have natural
generalizations to the case where $G$ is the group obtained by
exponentiation of a hyperbolic KM algebra. We refer readers
to \cite{Kac} for basic definitions and results of the theory of
KM algebras, and here only recall that a KM algebra ${\mathfrak{g}} \equiv 
{\mathfrak{g}} (A)$
is generally defined by means of a Cartan matrix $A$ and a set
of generators $\{e_i,f_i,h_i\}$ and relations (Chevalley-Serre
presentation), where $i,j=1,\dots, r\equiv {\rm rank}\, {\mathfrak{g}} (A)$.
The elements $\{ h_i \}$ span the Cartan subalgebra (CSA)
${\mathfrak{h}}$, while the $e_i$ and $f_i$
generate an infinite tower of raising and lowering operators, respectively.
The ``maximal compact'' subalgebra ${{\mathfrak{k}}}$ is defined as the 
subalgebra
of ${\mathfrak{g}} (A)$ left invariant under the Chevalley involution
$\omega (h_i) = -h_i\, , \, \omega (e_i)= -f_i\,,\, \omega (f_i)= -e_i$.
In other words, ${{\mathfrak{k}}}$ is spanned by the ``antisymmetric'' 
elements
$E_{\alpha,s} - E_{\alpha,s}^T $, where
$E_{\alpha,s}^T \equiv - \omega (E_{\alpha,s})$ is the ``transpose'' of
some multiple
commutator $E_{\alpha,s}$ of the $e_i$'s associated with the root $\alpha$ 
({\it i.e.}
$[ h, E_{\alpha,s}] = \alpha (h)  E_{\alpha,s}$ for $h\in {\mathfrak{h}}$). 
Here
$s=1,\dots {\rm mult} (\alpha)$ labels the different elements of 
${\mathfrak{g}} (A)$
having the same root $\alpha$. The (integer-valued) Cartan matrix of $E_{10}$ is 
encoded in its Dynkin diagram. See top diagram in Fig.~2.

The $\sigma$-model is formulated in terms of a one-parameter dependent group
element ${{\cal V}} ={{\cal V}}(t) \in \E$ and its Lie algebra valued 
derivative
\be
v(t) := \frac{d{{\cal V}}}{dt} {{\cal V}}^{-1} (t) \in {\mathfrak{e}}_{10} 
\equiv {\rm Lie} \, \E .
\ee
The action is
$\int dt {{\cal L}}$ with
\be\label{Lag0}
{{\cal L}} := {n(t)}^{-1} \langle v_{\rm sym}(t) | v_{\rm sym} (t)\rangle
\ee
with a ``lapse'' function $n(t)$ (not to be confused with $N$),
whose variation gives rise to the Hamiltonian constraint ensuring
that the trajectory is a null geodesic.
The ``symmetric'' projection $v_{\rm sym}:= \ft12 (v + v^T)$ eliminates
the component of $v$ corresponding to a displacement ``along 
${{\mathfrak{k}}}$'',
thereby defining an evolution on the coset space $\E/K(\E)$.
$\langle.|.\rangle$ is the standard invariant bilinear form on
the KM algebra \cite{Kac}.

Because no closed form construction exists for the raising operators
$E_{\alpha,s}$, nor their invariant scalar products
$\langle E_{\alpha,s} | E_{\beta,t} \rangle = N^\alpha_{s,t} \delta^0_{\alpha 
+ \beta}$, a recursive approach based on the decomposition
of $\E$ into irreducible representations of its ${{\rm SL} (10, {\Rn})}$ 
subgroup was devised in \cite{DHN2}. [See also \cite{KN1} for a decomposition of
$\E$ based on its $D_9$ subgroup, and \cite{KN2} for a decomposition of
$\E$ based on its $A_8 \times A_1$ subgroup.]
Let $\alpha_1,\dots,\alpha_9$ be the nine simple roots of $A_9 \equiv sl(10)$
corresponding to the horizontal line in the $\E$ Dynkin diagram,
and $\alpha_0$ the ``exceptional'' root connected to $\alpha_3$. [This root 
is 
labelled 10 in Fig.2.] Its dual
CSA element $h_0$ enlarges $A_9$ to the Lie algebra of ${\rm GL} (10)$.
Any positive root of $\E$ can be written as
\be\label{E10root}
\alpha = \ell \alpha_0 + \sum_{j=1}^9 m^j \alpha_j \quad (\ell,m^j \geq 0).
\ee
We call $\ell \equiv \ell (\alpha)$ the ``level'' of the root $\alpha$.
This definition differs from the usual one, where the (affine) level 
is identified with $m^9$ and thus counts the number of appearances 
of the over-extended root $\alpha_9$ in $\alpha$ \cite{FF,KMW}. Hence, 
our decomposition corresponds to a slicing (or ``grading'') of the
forward lightcone in the root lattice by spacelike hyperplanes, with only
finitely many roots in each slice, as opposed to the lightlike slicing
for the $E_9$ representations (involving not only infinitely many roots
but also infinitely many affine representations for $m^9\geq 2$ 
\cite{FF,KMW}).

The adjoint action of the $A_9$ subalgebra leaves the level $\ell(\alpha)$
invariant. The set of generators corresponding to a given level $\ell$
can therefore be decomposed into a (finite) number of irreducible
representations of $A_9$. The multiplicity of $\alpha$ as a root of
$\E$ is thus equal to the sum of its multiplicities as a weight
occurring in the ${{\rm SL} (10, {\Rn})}$ representations. Each irreducible 
representation
of $A_9$ can be characterized by its highest weight $\Lambda$, or
equivalently by its Dynkin labels $(p_1,\dots,p_9)$ where
$p_k (\Lambda) := (\alpha_k,\Lambda)\geq 0$ is the number of columns
with $k$ boxes in the associated Young tableau. For instance, the
Dynkin labels $(001 000 000)$ correspond to a Young tableau consisting
of one column with three boxes, {\it i.e.} the antisymmetric tensor
with three indices. The Dynkin labels are related to the 9-tuple of
integers $(m^1,\dots,m^9)$ appearing in \Ref{E10root} (for the
highest weight $\Lambda\equiv - \alpha$) by
\be\label{mi}
 S^{i3} \ell - \sum_{j=1}^9 S^{ij} p_j = m^i \geq 0
\ee
where $S^{ij}$ is the inverse Cartan matrix of $A_9$. This relation
strongly constrains the representations that can appear at level $\ell$,
because the entries of $S^{ij}$ are all positive, and the 9-tuples
$(p_1,\dots,p_9)$ and $(m_1,\dots, m_9)$ must both consist of
non-negative integers. In addition to satisfying the Diophantine
equations \Ref{mi}, the highest weights must be roots of $\E$, which
implies the inequality
\be\label{L2}
\Lambda^2 = \alpha^2 =
 \sum_{i,j=1}^9 p_i S^{ij} p_j - \ft1{10} \ell^2 \leq 2 .
\ee
All representations occurring at level $\ell +1$ are contained in
the product of the level-$\ell$ representations with the $\ell=1$
representation. Imposing the Diophantine inequalities \Ref{mi}, \Ref{L2} 
allows one to discard many representations appearing in this product.
The problem of finding a completely explicit and
manageable representation of $\E$ in terms of an infinite tower of
$A_9$ representations is thereby reduced to the problem of determining
the outer multiplicities of the surviving  $A_9$ representations,
namely the number of times each
representation appears at a given level $\ell$. The Dynkin labels
(all appearing with outer multiplicity one)
for the first six levels of $\E$ are
\begin{eqnarray}\label{irreps}
\ell=1  \quad &:& \quad (001000000) \non
\ell=2  \quad &:& \quad (000001000) \non
\ell=3  \quad &:& \quad (100000010) \non
\ell=4  \quad &:& \quad (001000001) \; , \; (200000000) \non
\ell=5  \quad &:& \quad (000001001) \; , \; (100100000) \non
\ell=6  \quad &:& \quad (100000011) \; , \; (010001000) \, ,\non
              &&            \quad (100000100) \; , \; (000000010) \, .
\end{eqnarray}
The level $\ell \leq 4$ representations can be easily determined by
comparison with the decomposition of $E_8$ under its $A_7$ subalgebra
and use of the Jacobi identity, which
eliminates the representations $(000000001)$ at level three and
$(010000000)$ at level four. By use of a computer and the $\E$
root multiplicities listed in \cite{KMW,BGN}, the
calculation can be carried much further \cite{Nicolai:2003fw}.

{}From
 \Ref{irreps} we can now directly read off the ${\rm} GL(10)$ tensors
making up the low level elements of $\E$. At level zero, we have the
$GL(10)$ generators ${K^a}_b$ obeying
$[{K^a}_b,{K^c}_d] = {K^a}_d \delta^c_b -  {K^c}_b \delta^a_d$. The 
${\mathfrak{e}}_{10}$
elements at levels $\ell=1,2,3$ are the $GL(10)$ tensors $E^{a_1a_2 a_3}$,
$E^{a_1 \dots a_6}$ and $E^{a_0|a_1 \dots a_8}$ with the symmetries
implied by the Dynkin labels.
The $\sigma$-model associates to these generators a corresponding tower
of ``fields'' (depending only on the ``time'' $t$): a zehnbein ${h^a}_b 
(t)$
at level zero, a three form $A_{abc} (t)$ at level one, a six-form
$A_{a_1\dots a_6} (t)$ at level two, a Young-tensor $A_{a_0|a_1 \dots 
a_8}(t)$
at level 3, etc. Writing the
generic $\E$ group element in Borel (triangular) gauge as
$ {{\cal V}}(t) = \exp X_h (t) \cdot \exp X_A (t)$
with $X_h(t) = {h^a}_b  {K^b}_a$ and $X_A (t) = \ft1{3!} A_{abc} E^{abc}
+ \ft1{6!} A_{a_1\dots a_6} E^{a_1\dots a_6} +
\ft1{9!} A_{a_0|a_1 \dots a_8} E^{a_0|a_1 \dots a_8} + \dots$, and using 
the
$\E$ commutation relations in $GL(10)$ form together with the bilinear
form for $\E$, we find up to third order in level
\begin{eqnarray}\label{Lag}
n {{\cal L}} &=& \ft14 (g^{ac} g^{bd} - g^{ab} g^{cd}) \dot g_{ab} \dot 
g_{cd}
  + \ft12 \ft1{3!} DA_{a_1a_2a_3}DA^{a_1a_2a_3} \non
&&  \!\!\!\!\!\!\!\! \!\!\!
 + \ft12 \ft1{6!} DA_{a_1 \dots a_6}DA^{a_1\dots a_6}
  + \ft12 \ft1{9!} DA_{a_0 | a_1 \dots a_8} DA^{a_0|a_1\dots a_8}
\end{eqnarray}
where $g^{ab} = {e^a}_c {e^b}_c$ with $ {e^a}_b \equiv {(\exp h)^a}_b$,
and all ``contravariant indices'' have been raised by $g^{ab}$. The
``covariant'' time derivatives are defined by (with $\partial A\equiv \dot 
A$)
\begin{eqnarray}\label{Dtime}
DA_{a_1a_2a_3} &:=& \partial A_{a_1a_2 a_3} \\
DA_{a_1\dots a_6} &:=& \partial A_{a_1 \dots a_6}
    + 10 A_{[a_1a_2 a_3} \partial A_{a_4a_5 a_6]} \non
DA_{a_1|a_2\dots a_9} &:=& \partial A_{a_1|a_2 \dots a_9}
    + 42 A_{\langle a_1a_2 a_3} \partial A_{a_4 \dots  a_9 \rangle} \non
&&  \!\!\!\! \!\!\!\!\!\!\!\! \!\!\!\!\!\!\!\! \!\!\!\!\!\!\!\! 
\!\!\!\!\!\!
- 42 \partial A_{\langle a_1a_2 a_3} A_{a_4 \dots  a_9 \rangle}
    + 280 A_{\langle a_1a_2 a_3} A_{a_4a_5 a_6} \partial A_{a_7a_8 
a_9\rangle} .
\nonumber
\end{eqnarray}
Here antisymmetrization $[\dots]$, and projection on the $\ell = 3$
representation $\langle \dots \rangle$, are  normalized with strength one
({\it e.g.} $[[\dots]] = [\dots]$). Modulo field redefinitions,
all numerical coefficients in \Ref{Lag} and \Ref{Dtime} are uniquely
fixed by the structure of $\E$. See also \cite{DN04} for a different perspective
on the equations of motion of the $E_{10} / K (E_{10})$ $\s$-model.

The Lagrangian \Ref{Lag0} is invariant under a nonlinear
realization of $\E$ such that ${{\cal V}}(t) \rightarrow k_g(t) {{\cal V}}(t) 
g$ 
with
$g\in\E$; the compensating ``rotation'' $k_g(t)$  being, in general,
required to restore
the ``triangular gauge''. When $g$ belongs to the nilpotent
subgroup generated by the $E^{abc}$, etc., this symmetry reduces to
the rather obvious ``shift'' symmetries of \Ref{Lag}
and no compensating rotation is needed. The latter are, 
however, required for the transformations generated by 
$F_{abc} = (E^{abc})^T$, etc. The associated infinite
number of conserved (Noether) charges are formally 
given by $J={{\cal M}}^{-1} \partial {{\cal M}}$, where ${{\cal M}} \equiv 
{{{\cal V}}}^T {{\cal V}}$. 
This can be formally solved in closed form as
\be\label{M}
{{\cal M}} (t) = {{\cal M}} (0) \cdot \exp (tJ).
\ee
The compatibility between \Ref{M} (indicative of the integrability 
of \Ref{Lag}) and the chaotic behavior of $g_{ab}(t)$ near a 
spacelike singularity is discussed in \cite{Damour:2002et}.

The main result of concern here is the following:
there exists a {\it map} between geometrical quantities constructed
at a given spatial point $x$ from the supergravity fields
$G_{\mu\nu}(x^0,x)$ and $\cA_{\mu\nu\rho}(x^0,x)$ and the
one-parameter-dependent quantities $g_{ab}(t), A_{abc} (t), \dots$ 
entering the coset Lagrangian \Ref{Lag}, under which the supergravity 
equations of motion \Ref{EOM} become {\it equivalent, up to 30th order in 
height}, to the Euler-Lagrange equations of \Ref{Lag}.
In the gauge \Ref{Gauge} this map is defined by
$t = x^0  \equiv \int dT/ \sqrt{G}$ and
\begin{eqnarray}\label{map}
g_{ab}(t) &=& G_{ab} (t,x) \\
DA_{a_1a_2a_3}(t)  &=& {{\cal F}}_{0a_1 a_2 a_3} (t,x) \non
DA^{a_1 \dots a_6} (t) &=&   - \ft1{4!}
\varepsilon^{a_1\dots a_6 b_1 b_2 b_3 b_4} {{\cal F}}_{b_1 b_2 b_3 b_4} (t,x) 
\non
DA^{b|a_1 \dots a_8 } (t)&=& \ft32 \varepsilon^{a_1\dots a_8 b_1 b_2}
\big( {C^b}_{b_1 b_2} (x) + \ft29 \delta^b_{[b_1}  {C^c}_{b_2] c} (x) \big).
\nonumber
\end{eqnarray}
The expansion in height ${\rm ht} (\alpha) \equiv \ell + \sum m^j$, which
controls the iterative validity of this equivalence, is as follows:
the Hamiltonian constraint of the coset model \Ref{Lag} contains an
infinite series of exponential coefficients $\exp\big(-2\alpha(\beta)\big)$,
where $\alpha$ runs over all positive roots of $\E$, and where
$\beta^a\equiv - {h^a}_a$ parametrize the CSA of $\E$. The {\it billiard 
picture} 
discussed above is equivalent to saying that, near a spacelike singularity 
($t 
\rightarrow \infty$),
the dynamics of the
supergravity fields and of truncated versions of the $\E$ coset
fields is asymptotically dominated by the (hyperbolic) Toda model
defined by keeping only the exponentials involving the {\it simple
roots} of $\E$. Higher roots introduce smaller and smaller
corrections as $t$ increases. The ``{\it height expansion}'' of 
the equations of motion is then technically defined as a formal
BKL-like expansion that corresponds to such an expansion in
decreasing exponentials of the Hamiltonian constraint. On the
supergravity side, this expansion amounts to an expansion in 
gradients of the fields in appropriate frames. Level one 
corresponds to the simplest one-dimensional reduction of
\Ref{EOM}, obtained by assuming that both $G_{\mu \nu}$ and
$\cA_{\lambda \mu \nu}$ depend only on time \cite{DH2}; levels 2 and
3 correspond to configurations of $G_{\mu \nu}$ and $\cA_{\lambda \mu \nu}$ 
with a more general, but still very restricted $x$-dependence, 
so that {\it e.g.} the frame derivatives of the electromagnetic field 
in (\ref{EOM}) drop out \cite{DHHS}. In \cite{DHN2,DN04} it is checked that, when
neglecting terms
corresponding to ${\rm ht} (\alpha)\geq 30$, the map \Ref{map} 
provides a {\it perfect match} between the supergravity evolution 
equations \Ref{EOM} and the $\E$ coset ones, as well as
between the associated Hamiltonian constraints. (In fact, the 
matching extends to {\it all real roots} of level $\leq 3$.)
Let us also mention in passing (from \cite{DN04})
that the $\E$ coset action is not compatible with the addition of an
eleven-dimensional cosmological constant in the supergravity action
(an addition which has been proven to be incompatible with supersymmetry in
\cite{BDHS}).

It is natural to view the map (\ref{map}) as embedded in a hierarchical 
sequence of maps involving more and more spatial gradients 
of the basic supergravity fields. The height expansion 
would then be a way of revealing step by step a hidden 
hyperbolic symmetry, implying the existence of a 
huge non-local symmetry of Einstein's theory and its
generalizations. Although the validity of this conjecture remains
to be established, one can at least show that there is ``enough
room'' in $\E$ for all the spatial gradients. Namely, the search for
affine roots (with $m^9 =0$) in \Ref{mi} and \Ref{L2} reveals 
three infinite sets of admissible $A_9$ Dynkin labels
$(00100000n), (00000100n)$ and $(10000001n)$ with highest
weights obeying $\Lambda^2 =2$, at levels $\ell=3n+1,3n+2$ and $3n+3$, 
respectively. These correspond to three infinite towers of $\e$ elements 
\be\label{affine} 
{E_{a_1\dots a_n}}^{b_1 b_2b_3} \; , \;
{E_{a_1\dots a_n}}^{b_1 \dots b_6} \; , \; {E_{a_1\dots
a_n}}^{b_0| b_1 \dots b_8} 
\ee
which are symmetric in the lower indices and all appear with 
outer multiplicity one (together with three transposed
towers). Restricting the indices to $a_i=1$ and $b_i \in
\{2,...,10 \}$  and using the decomposition ${\bf 248}
\!\rightarrow\! {\bf 80}\!+ \!{\bf 84}\! +\! \overline{\bf 84}$ of
$E_8$ under its ${\rm SL}(9)$ subgroup one easily recovers the
affine subalgebra $E_9\subset\E$. The appearance of higher order
dual potentials ({\it \`a la} Geroch) in the $E_9$-based linear
system for $D\!=\!2$ supergravity \cite{BM} indeed suggests that
we associate the $\E$ Lie algebra elements \Ref{affine} to the
higher order spatial gradients $\partial^{a_1} \cdots \partial^{a_n} A_{b_1
b_2 b_3}, \partial^{a_1} \cdots \partial^{a_n} A_{b_1 \dots b_6}$ and
$\partial^{a_1} \cdots \partial^{a_n} A_{b_0|b_1 \dots b_8}$ or to some of
their non-local equivalents. Finally, we refer to \cite{Damour:2002et} for a 
more ge\-neral discussion of the height expansion of Kac-Moody 
$\sigma$-models, 
and we note that the approach of \cite{DHN2,Damour:2002et} can be extended to 
other physically relevant KM
algebras, such as $BE_{10}$ \cite{DH3,CJLP3} and $AE_n$
\cite{DHJN}.

\end{document}